\documentclass[aps,prx,twocolumn,letterpaper,superscriptaddress,notitlepage,longbibliography,nofootinbib]{revtex4-1}
\usepackage{xcolor,graphicx,amsmath,amsfonts,amssymb,amsthm,bm,bbm}
\usepackage{tikz,pgfplots}
\pgfplotsset{compat=newest}
\pgfplotsset{plot coordinates/math parser=false}
\usetikzlibrary{arrows,decorations.pathmorphing,backgrounds,positioning,fit,calc,patterns,shapes,decorations.markings,shadings}
\usetikzlibrary{external}
\usepackage[pdfpagelabels,pdftex,bookmarks,breaklinks]{hyperref}
\definecolor{darkblue}{RGB}{0,0,127} % choose colors
\definecolor{darkgreen}{RGB}{0,180,0}
\hypersetup{
	colorlinks, 
	linkcolor=darkblue, 
	citecolor=darkgreen, 
	filecolor=red, 
	urlcolor=blue,
	pdftitle={Detecting Topological Order with Ribbon Operators}, 
	pdfauthor={Jacob C. Bridgeman, Steven T. Flammia, David Poulin}
}

\newcommand{\ZZ}{\mathbb{Z}}

\newcommand{\Eref}[1]{Eq.~(\ref{#1})}
\newcommand{\Sref}[1]{Sec.~\ref{#1}}
\newcommand{\Fref}[1]{Fig.~\ref{#1}}

\let\oldonlinecite\onlinecite
\renewcommand{\onlinecite}[1]{Ref.~[\oldonlinecite{#1}]}

\newcommand{\cO}{\mathcal{O}}
\DeclareMathOperator{\Tr}{Tr}

\newcommand{\ket}[1]{|{#1}\rangle}

\newcommand{\bra}[1]{\langle{#1}|}

\newcommand{\braket}[2]{\langle{#1}|{#2}\rangle}

\newcommand{\norm}[1]{\left\lVert#1\right\rVert}
\newcommand{\opnorm}[1]{\left\lVert#1\right\rVert_\text{op}}
\newcommand{\comm}[1]{\left[#1\right]}
\newcommand{\restrict}[1]{\raise-.2ex\hbox{\ensuremath|}_{#1}}

\newcommand{\midarrow}{\tikz \draw[-stealth] (0,0) -- +(.1,0);}

\newcommand{\MPOtensor}[3]{
\def\a{#1}
\def\dx{#2}
\def\x{#3}
\draw[shift={(\x*\a+\x*\dx,0)}] (-\dx/2,0) -- (0,0);
\filldraw[a,shift={(\x*\a+\x*\dx,0)}] (0,0) -- (\a/2,\a/2) -- (\a,0) -- (\a/2,-\a/2) -- (0,0);
\draw[shift={(\x*\a+\x*\dx,0)}] (\a,0) -- (\a+\dx/2,0);
\draw[shift={(\x*\a+\x*\dx,0)}] (\a/2,\a/2) -- node {\midarrow}(\a/2,\a/2+\dx/2);
\draw[shift={(\x*\a+\x*\dx,0)}] (\a/2,-\a/2-\dx/2) -- node {\midarrow} (\a/2,-\a/2);
}

\makeatletter
\newcommand{\vast}{\bBigg@{4}}
\newcommand{\Vast}{\bBigg@{9}}
\makeatother

\def\Put(#1,#2)#3{\leavevmode\makebox(0,0){\put(#1,#2){#3}}}

\makeatletter
\def\pgf@plot@curveto@handler@finish{%
  \ifpgf@plot@started%
    \pgfpathcurvebetweentimecontinue{0}{0.995}{\pgf@plot@curveto@first}{\pgf@plot@curveto@first@support}{\pgf@plot@curveto@second}{\pgf@plot@curveto@second}%
  \fi%
}
\makeatother

\newlength\figureheight 
\newlength\figurewidth 
  
\newcommand{\includeTikz}[2]{\includegraphics{#1.pdf}
}
\begin{document}

\title{Detecting Topological Order with Ribbon Operators}

\author{Jacob C.\ Bridgeman}
\author{Steven T.\ Flammia}
\affiliation{Centre for Engineered Quantum Systems, School of Physics, The University of Sydney, Sydney, Australia}
\author{David Poulin}
\affiliation{Department of Physics, Universit\'{e} de Sherbrooke, Sherbrooke, Qu\'{e}bec, Canada}

\date{\today}

\begin{abstract}
We introduce a numerical method for identifying topological order in two-dimensional models based on one-dimensional bulk operators. The idea is to identify approximate symmetries supported on thin strips through the bulk that behave as string operators associated to an anyon model. We can express these ribbon operators in matrix product form and define a cost function that allows us to efficiently optimize over this ansatz class. We test this method on spin models with abelian topological order by finding ribbon operators for $\mathbb{Z}_d$ quantum double models with local fields and Ising-like terms. In addition, we identify ribbons in the abelian phase of Kitaev's honeycomb model which serve as the logical operators of the encoded qubit for the quantum error-correcting code. We further identify the topologically encoded qubit in the quantum compass model, and show that despite this qubit, the model does not support topological order. Finally, we discuss how the method supports generalizations for detecting nonabelian topological order. 
\end{abstract}

\maketitle

Despite the apparent simplicity of quantum spin models, they can exhibit a wide variety of interesting and potentially useful phenomena. These range from conventional magnetic order to the more novel topological\cite{Wen2013} and symmetry-protected\cite{Chen2013c} and symmetry-enriched\cite{Barkeshli2014} topological orders which are of interest in both condensed matter physics\cite{Anderson} and quantum information theory.\cite{Kitaev2003} These states are disordered in the sense of Landau-Ginzburg-Wilson, however they do exhibit properties distinct from the usual disordered phases. For example, topological phases possess quasiparticle excitations, known as anyons, whose braid relations can be far more exotic than those of fermions or bosons.\cite{Nayak2008} These phases also have ground state degeneracy which depends on the topology of the lattice.\cite{Wen1990} This protected degeneracy has prompted the investigation of topologically ordered models as quantum memories.\cite{Kitaev2003, Dennis2002} Quantum information stored in the degenerate subspace can be protected from arbitrary local noise when error correction techniques are employed.\cite{Brown2014,Terhal2015,Brell2014} 

Distinguishing topological phases can be an especially challenging task precisely because of their topological nature: there is no broken symmetry and no local order parameter signalling the phase transition.\cite{Wen2013} There has been a large amount of previous work which attempts to identify topological order (TO). The existing key techniques, such as the topological entanglement entropy,\cite{Levin2006, Kitaev2006a, Flammia2009b, Isakov2011, Zhang2012, Jiang2012} the entanglement Hamiltonian\cite{Li2008}, topological degeneracy\cite{Trebst2007} and associated properties,\cite{Zhang2012, Cincio2013} symmetries of particular representations of the ground states,\cite{Schuch2010, Buerschaper2014, Sahinoglu2014, Bultinck2015, Liu2015} and specific properties of particular TO states,\cite{Evenbly2010, Yan2011} have been highly successful in various domains of applicability. We review these methods below. A common feature of these methods is that they utilize the ground state of the model, which is unfortunately a challenging computational task in general. 

Here we propose a numerical method that we call the \emph{ribbon operators} method for identifying TO in the ground state of a given 2D Hamiltonian using only the Hamiltonian, without reference to the ground state. We reduce the search for TO to a 1D problem through the bulk of the material, and we present a variational approach based on standard DMRG\cite{Schollwock2011} to identify certain operators -- the ribbon operators -- supported in the bulk that satisfy the commutation relations relevant for a candidate anyon model. We demonstrate the power of this approach by identifying TO in both integrable and non-integrable models, and contrast this with topologically trivial Hamiltonians. We also demonstrate the ability to identify topologically encoded qubits and logical operators of quantum error-correcting codes, even in a non-integrable model. All of our calculations are focused on the case of abelian TO in spin models, however the ribbon operators method suggests several natural extensions beyond this case, which we leave open for future work.

The following subsection reviews prior approaches, while \Sref{S:anyons} reviews anyon models. The expert reader can skip to \Sref{S:ribbons}.

\subsection{Prior Approaches}

One important tool is the \emph{topological entanglement entropy} (TEE)\cite{Levin2006, Kitaev2006a} of the ground state wave function. Given the reduced density matrix $\rho_R$ of some many-body ground state on a region $R$, the von Neumann entropy $S(\rho_R)=-\Tr\left(\rho_R\log\rho_R\right)$ typically obeys the area law
\begin{align}
  S(\rho_R)&=\alpha |\partial R|-\gamma+\cO\left(\frac{1}{|R|}\right),\label{eqn:arealaw}
\end{align}
where $|R|$ and $|\partial R|$ are the number of spins in the region $R$ and on the boundary of the region $R$ respectively, and $-\gamma$ is the TEE.

In a topologically ordered model, a physical argument suggests that $\gamma=\log\left(\sqrt{\sum_c d_c^2}\right)$, where $d_c$ is the quantum dimension of the anyon with charge $c$ in the associated anyon model. The TEE is clearly nonzero if the ground state is topologically ordered,\cite{Levin2006} so $\gamma$ can be used as a signal of such ordering.\cite{Isakov2011, Zhang2012, Jiang2012}  
One can compute $\gamma$ by obtaining an explicit ground state wavefunction, for example as a PEPS. Additionally, a R\'{e}nyi entropy variant of $\gamma$\cite{Flammia2009b} can be computed by sampling the ground state wavefunctions using quantum Monte Carlo.\cite{Isakov2011}

There are two major challenges associated with this approach. Firstly, there exist examples of topologically trivial states for which computing $\gamma$ leads to nonzero values.\cite{Zou} Secondly, any two topological phases whose associated anyon models have the same total quantum dimension $\mathcal{D}=\sum_cd_c^2$ will have the same TEE, and so it cannot be used to distinguish them. This second problem has already led to difficulty in fully identifying the phase of physically interesting models including the Heisenberg antiferromagnet on the Kagome lattice, a model thought to describe the low-energy physics of several naturally occurring and synthetic minerals.\cite{Jiang2012}

A less coarse approach is to investigate the full entanglement spectrum, that is, the spectrum of the effective Hamiltonian defined by $\rho_R=\mathrm{e}^{-H_{\mathrm{eff}}}$.\cite{Li2008} It has been suggested that the universal properties of $H_\mathrm{eff}$ are intimately linked to the structure of the physical edge state of the model, however $H_\mathrm{eff}$ can undergo phase transitions without the physical model doing so,\cite{Chandran2014} and at least for models in the same phase as a string-net model\cite{Levin2005} the spectrum is expected to contain the same universal information as the TEE.\cite{Flammia2009b}

One of the characteristic features of topologically ordered models is the topology-dependent ground state degeneracy.\cite{Wen1990} After obtaining a full set of ground states, one can observe transitions out of topological phases via loss of topological degeneracy,\cite{Trebst2007} and, in the topological phase, compute the $\mathcal{S}$ and $\mathcal{U}$ matrices defining the braiding relations in the associated anyon theory.\cite{Zhang2012, Cincio2013} Unfortunately, demonstrating topology-dependent ground state degeneracy is not sufficient to ensure robust topological order.\cite{Nussinov2006, Nussinov2007}

Given a projected entangled pair state (PEPS) description of the ground state, one can identify the topological order by understanding the symmetry properties of the parent Hamiltonian\cite{Schuch2010} or the environment tensor.\cite{Liu2015} One can also use the PEPS formalism to identify matrix product operators (MPOs) that `pull through' the PEPS tensors on the virtual level.\cite{Sahinoglu2014, Bultinck2015} These MPOs are in close analogy to the physical operators that act to create and transport anyons. As we will describe, these physical operators are central to our approach.  

Finally, given access to a ground state, specific structure such as certain correlation functions or distribution of bond energies can provide evidence for topological ordering.\cite{Evenbly2010, Yan2011}

As a prerequisite for each of these methods, one must obtain an efficient description of the ground space (e.g.\ via tensor networks) or obtain expectations with respect to ground states (e.g.\ via Monte Carlo sampling). Tensor network (TN) methods have proven very useful for this purpose,\cite{Jiang2012, Cirac2011, Yan2011, Evenbly2010, Liu2015} however in many cases the result does not conclusively determine the topological order.\cite{Yan2011} The TN states obtained suffer from several drawbacks. In particular, they are usually computed on infinite cylinders of small circumference. They can also be biased towards low-entanglement states.\cite{Evenbly2010} Additionally, properties of a ground state alone are not sufficient to identify a gapped topological phase. For example, a one can obtain a gapless Hamiltonian sharing the toric code ground space.\cite{Freedman2008}

Of these methods, the ones closest in spirit to our current approach are the tensor network-based methods,\cite{Schuch2010, Sahinoglu2014, Bultinck2015} however our approach differs substantially in that we do not require a PEPS description of the ground state wavefunction. We instead variationally create tensor network representations of certain ribbon operators that are supported on 1D strips through the truly two-dimensional bulk; this dimensional reduction is what makes our method numerically tractable. In contrast to state-based variational approaches, where a similar reduction is often included for numerical convenience, we will argue that in our operator-based approach this reduction is an expected feature of the operators. Additionally, if one wishes to use the topologically protected ground space as a quantum memory, knowledge of these ribbon operators is required for information manipulation and extraction. Thus, this method could be used to augment a state-based approach to obtain this additional data.

This paper is organized as follows. In \Sref{S:anyons}, we review some of the features of anyon models which describe the low energy excitations of topologically ordered spin models. In \Sref{S:ribbons}, we define a ribbon operator, and use the properties of the anyons to define a cost function which quantifies how well a candidate ribbon realizes their behavior. We then describe how the cost function can be minimized numerically in \Sref{S:algorithm}. Using this algorithm, in \Sref{S:Results} we obtain ribbon operators in a number of topologically ordered and topologically trivial models. These numerics demonstrate that the cost function we define allows identification of abelian topological order in nonintegrable spin models. 
In \Sref{S:LogicalOperators}, we prove that, under certain assumptions, ribbon operators can be used as approximate logical operators in quantum error correcting codes. We conclude in \Sref{S:future} with a discussion of extensions of the method to nonabelian topological order and more complex spin models.

\section{Properties of Anyons}\label{S:anyons}
%------------------------------------------------------------------------------------------------------------%
\begin{figure}
  \centering
  \includeTikz{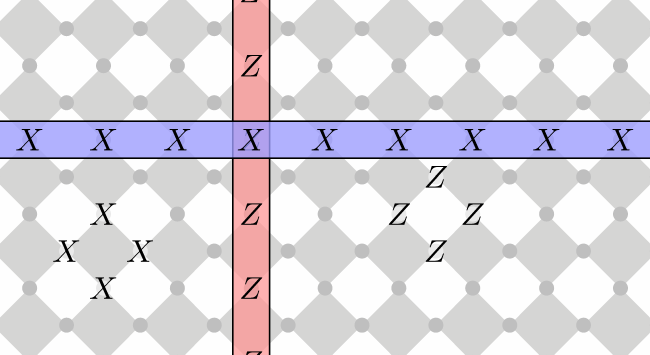}{
  \tikzsetnextfilename{toricStrings}
  \begin{tikzpicture}
    [a/.style={fill=black!20},
        b/.style={fill=white},
        c/.style={fill=red,fill opacity = .3},
        d/.style={fill=blue,fill opacity = .3},
        e/.style={fill=black!30},
        scale=.75]
    \def\dx{1}
    \def\dy{1};
    \def\maxx{10};
    \def\maxy{8};
    \def\r{.1};
    \clip (.6,.6) rectangle (9.4,5.4);
    \foreach \x in {1,2,...,\maxx}
    {
    \foreach \y in {1,2,...,\maxy}
    {
    \fill[b,shift={({(\x-.5)*\dx},{(\y-.5)*\dy})}] (-\dx/2,0)--(0,\dy/2)--(\dx/2,0)--(0,-\dy/2)--(-\dx/2,0);
    }
    }
    \foreach \x in {0,1,...,\maxx}
    {
    \foreach \y in {0,1,...,\maxy}
    {
    \fill[a,shift={(\x*\dx,\y*\dy)}] (-\dx/2,0) {} --(0,\dy/2)--(\dx/2,0)--(0,-\dy/2)--(-\dx/2,0);
    \fill[e,shift={(\x,\y)}] (-\dx/2,0) circle (\r);
    \fill[e,shift={(\x,\y)}] (0,-\dy/2) circle (\r);
    }
    };
    \fill[fill=white,opacity=.6] (3.75,.5)--(4.25,.5)--(4.25,6.5)--(3.75,6.5)--(3.75,.5);
    \filldraw[c] (3.75,.5)--(4.25,.5)--(4.25,6.5)--(3.75,6.5)--(3.75,.5);
    \draw[shift={(4,3)}] (0,\dy/2) node {$Z$};
    \filldraw[fill=white,opacity=.7,shift={(0,.5)}]
     (.1,2.75)--(10.5,2.75)--(10.5,3.25)--(.1,3.25)--(.1,2.75);
        \filldraw[d,shift={(0,.5)}] (.1,2.75)--(10.5,2.75)--(10.5,3.25)--(.1,3.25)--(.1,2.75);
    \foreach \y in {-1,0,...,3}{\draw[shift={(\y,3)}] (0,\dy/2) node {$X$};};
	\foreach \y in {5,6,...,10}{\draw[shift={(\y,3)}] (0,\dy/2) node {$X$};};
	\foreach \y in {-1,0,...,2}{\draw[shift={(4,\y)}] (0,\dy/2) node {$Z$};};
	\foreach \y in {4,5,...,10}{\draw[shift={(4,\y)}] (0,\dy/2) node {$Z$};};
\draw[shift={(4,3)}] (0,\dy/2) node {$X$};
   \fill[black!10,shift={(2,2)}] (-\dx/2,0) circle (\r);
   \fill[black!10,shift={(2,2)}] (\dx/2,0) circle (\r);
   \fill[black!10,shift={(2,2)}] (0,\dy/2) circle (\r);
   \fill[black!10,shift={(2,2)}] (0,-\dy/2) circle (\r);
      \fill[black!10,shift={(6.5,2.5)}] (-\dx/2,0) circle (\r);
      \fill[black!10,shift={(6.5,2.5)}] (\dx/2,0) circle (\r);
      \fill[black!10,shift={(6.5,2.5)}] (0,\dy/2) circle (\r);
      \fill[black!10,shift={(6.5,2.5)}] (0,-\dy/2) circle (\r);
   \path[shift={(2,2)}] (-\dx/2,0)node {$X$}--(0,\dy/2)node {$X$}--(\dx/2,0)node {$X$}--(0,-\dy/2)node {$X$}--(-\dx/2,0);
   \path[shift={(6.5,2.5)}] (-\dx/2,0)node {$Z$}--(0,\dy/2)node {$Z$}--(\dx/2,0)node {$Z$}--(0,-\dy/2)node {$Z$}--(-\dx/2,0);
  \end{tikzpicture}}
  \caption{The prototypical ribbon operators are the string operators shown above for the toric code. The toric code is defined by $X$-type and $Z$-type four-body interactions on alternate plaquettes of a square lattice, and the (exact) ribbon operators are the string-like products of $X$ and $Z$ shown above. These ribbon operators commute with the Hamiltonian terms but anticommute with each other, thus identifying a phase with $\ZZ_2$ topological order.}\label{fig:toricStrings}
%  \vspace*{-5mm}
\end{figure}
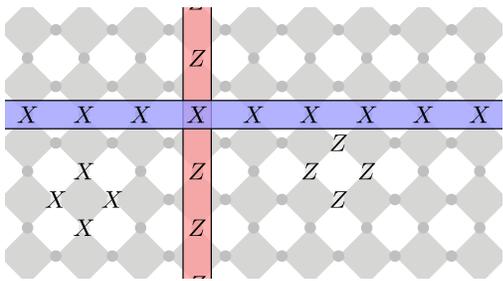

In this section, we review the physical properties of anyon models that characterize a topological phase. Of course, not all topological phases have an associated anyon model, for example the cubic code,\cite{Haah2011} however we will tailor our method to those phases with anyonic excitations. This discussion of anyon models will motivate our definition in the subsequent section of a \emph{ribbon operator} for a spin model defined by some local Hamiltonian $H=\sum_j h_j$ on a lattice in two dimensions. We will argue that this definition captures the essential features outlined below of a topologically ordered model by thinking about anyons as quasiparticle excitations of the spin model. The simplest realization of this is Kitaev's toric code,\cite{Kitaev2003, Kitaev2006} and we will refer to that model to clarify key features.

Creating a pair of anyons, and moving them to the boundary of the lattice leaves no excitations, and so maps the vacuum to the vacuum. In the toric code, anyons are created and transported using string operators, as shown in \Fref{fig:toricStrings}. These operators commute with the Hamiltonian in the bulk, but have excitations localized to the ends. In this way, if the anyons are moved off the edge of the lattice or are fused together, the system remains in its ground state.

More generally, we expect anyonic excitations to be particle like, but to have nontrivial braiding relations. On sufficiently large length scales, it is expected that these braid statistics are governed solely by an anyon model.\cite{Kitaev2006} For this reason, the creation operators in this limit should be one dimensional with the particles localized around the end points. Away from a renormalization-group fixed point, we expect these operators to be dressed, and supported on slightly fattened regions, though still effectively one-dimensional. More precisely, the width of these ribbons should be set by the microscopic details of the model, and should not scale with system size.

The family of local commuting projector codes (LCPC) generalizes the toric code and provides a concrete instantiation of these ideas. The LCPCs are families of Hamiltonians where the terms are pairwise commuting projectors onto an unfrustrated ground state. In these models the string operators described above always exist.\cite{Haah2012} This class of models includes exactly solvable points of most known topological phases\cite{Landon-Cardinal2013} including the Levin-Wen string nets.\cite{Levin2005} Away from these exactly solvable points, we can dress the string operators using quasiadiabatic continuation.\cite{Hastings2005} Given a point in a phase for which an initial string operator exists, such as an LCPC, we can dress this operator to any other point in the phase. Generically this operator will have extensive support, however the error made in truncating the width is exponentially small in $w/\xi$, where $w$ is the width and $\xi$ is the correlation length. Thus string-like operators with width proportional to the correlation length can be found anywhere within the phase.

Since the string operators act as creation operators for pairs of anyons, we argue that they should be only slightly entangling operators.\cite{Haah2012} These are operators whose Schmidt rank is independent of the system size on any bipartition that cuts the string into two contiguous pieces. In the LCPC case, the string operators are known to be slightly entangling.\cite{Haah2012} Since the quasiparticles are simply dressed by moving away from these fixed points, we expect this property to remain throughout the phase; although the required Schmidt rank can increase substantially, it will still be independent of the system size. When we remove the dressed quasiparticles from the lattice (by annihilating them or bringing them out to infinity), the system returns to its ground space, which is expected to obey the area law of \Eref{eqn:arealaw}. This also supports the slightly entangling nature of such operators.

Some of the defining features of an anyon model are the relations that are encoded in the topological $\mathcal{S}$ matrix and the $\mathcal{R}$ matrices. Let $\{a,b,c,\ldots\}$ label the anyonic particles in the anyon model describing the low energy excitations of the Hamiltonian $H$. 
The $\mathcal{R}$ matrices are defined by
\begin{align}
\begin{array}{c}
\includeTikz{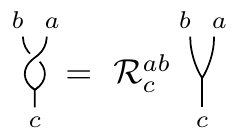}{
    \tikzsetnextfilename{Rmatrix}
\begin{tikzpicture}
	[a/.style={draw=white, double=black, double distance = .5pt, scale=.6, line width = 1.5pt},
	b/.style={line width=.5pt, scale=.6,rounded corners = .6em}],
	\draw[a] {[rounded corners = .75em] (0,.5) -- (.38,.75)} {[rounded corners = .3em] -- (-.2,1.2)} -- (-.2,1.4);
	\draw[a] {[rounded corners = .75em] (0,.5) -- (-.38,.75)} {[rounded corners = .3em]  -- (.2,1.2)} -- (.2,1.4);
	\draw[line width=.5, scale=.6] (0,.2) -- (0,.5) -- (-.067,.55) (0,.5) -- (.067,.55);	
	\node[anchor=north] at (0,.15) {$\scriptstyle c$};
	\node[anchor=south] at (-.175,.8) {$\scriptstyle b$};
	\node[anchor=south] at (.175,.8) {$\scriptstyle a$};
\node[anchor=south] at (.85,.15) {$=\ \mathcal{R}_c^{ab}$};
\begin{scope}[shift={(1.7,0)}]
	\draw[b] (0,.7) -- (.2,1)  -- (.2,1.4);
	\draw[b] (0,.7) -- (-.2,1)  -- (-.2,1.4);
	\draw[b] (0,.2) -- (0,.7);
	\node[anchor=north] at (0,.15) {$\scriptstyle c$};
	\node[anchor=south] at (-.175,.8) {$\scriptstyle b$};
	\node[anchor=south] at (.175,.8) {$\scriptstyle a$};
\end{scope}
\end{tikzpicture}}
\end{array}\!,\label{eqn:Rmatrix}
\end{align}
and the $\mathcal{S}$ matrix is 
\begin{align}
\mathcal{S}_{ab} = \frac{1}{\mathcal{D}}\!\!
\begin{array}{c}
\includeTikz{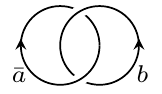}{
\tikzsetnextfilename{Smatrix}
\begin{tikzpicture}
	[a/.style={draw=white, double=black, double distance = .5pt, scale=.4, line width = 1.5pt},b/.style={arrows={-{stealth}},scale=.4}]
	\draw[a] (0,0) arc [start angle=0, end angle=180, radius=1];
	\draw[a] (1,0) arc [start angle=0, end angle=360, radius=1];
	\draw[a] (0,0) arc [start angle=0, end angle=-180, radius=1];
	\draw[b] (-2,0.15)--(-2,0.17);
	\draw[b] (1,0.15)--(1,0.17);
	\node[anchor=south east] at (-.6,-.5) {${}_{\bar{a}}$};
	\node[anchor=south west] at (.25,-.5) {${}_b$};
\end{tikzpicture}}
\end{array}
\!,\label{eqn:Smatrix}
\end{align}
where the lines trace out the worldlines of the particles. 

Another approach is to consider the related $\tilde{\mathcal{S}}$ matrix that can be obtained via the twist product.\cite{Haah2014} If the $\mathcal{S}$ matrix is an object that captures dynamical information from braiding anyons in 2+1 dimensions, then the $\tilde{\mathcal{S}}$ matrix is a static object that captures correlations of ground states in 2+0 dimensions. If $\pi_a$ is an operator supported on an annulus that projects onto a state with total charge $a$ inside, then
\begin{align}\label{eqn:twistS}
  \tilde{\mathcal{S}}_{ab}&=\bra{\psi}\pi_a\infty \pi_b\ket{\psi} = \frac{d_a d_b}{\mathcal{D}} \mathcal{S}_{ab}\,,
\end{align}
where $\ket{\psi}$ is any ground state of the spin model. For bipartite operators $X=\sum_{i,j}X_{i}^{A}\otimes X_{j}^{B}$ and $Y=\sum_{k,l}Y_{k}^{A}\otimes Y_{l}^{B}$, the twist product with respect to the partition $A|B$ is defined as
\begin{align}
  X\infty Y&=\sum_{i,j,k,l}X_i^{A}Y_k^{A}\otimes Y_l^{B}X_j^{B},
\end{align}
where the order is reversed on the $B$ region relative to the $A$ partition. This is clearly closely related to \Eref{eqn:Smatrix}, where the worldline loops are replaced by the corresponding closed loop operators on the lattice. For abelian models where $d_a =1$ for all particle types $a$, the $\mathcal{S}$ and $\tilde{\mathcal{S}}$ matrices coincide up to a multiple of $\mathcal{D}$. 
  
Finally, anyonic quasiparticles should be able to move around the lattice, and the state should not depend upon any smooth deformations of the path they take. In particular, when they are fused back together or moved off the lattice, the system should return to a vacuum state. This is realized in the toric code since the string operators can be deformed by dressing with local operators that are symmetries of the Hamiltonian. This preserves commutativity with $H$ and the $\tilde{\mathcal{S}}$ matrix and clearly creates the same particle at the ends.
  
\section{Definition of Ribbon Operators}\label{S:ribbons}
%------------------------------------------------------------------------------------------------------------%

Let us summarize the physical intuition that we've gained in the previous section into a few simple properties that the string-like operators seem to possess in general. These properties hold for known exactly solvable models with TO. 

Given a two-dimensional quantum system with TO, we expect that there are operators $R$ supported on one-dimensional strips through the bulk where the following four properties hold, at least approximately:
\begin{enumerate}
	\item $R$ commutes with the bulk Hamiltonian in the low energy sector.\label{item:commuting}
	\item $R$ is supported on a strip of spins with bounded width $w$.\label{item:bounded}
	\item Distinct ribbons $R$ and $L$ should respect the data (e.g.\ the $\mathcal{R}$ or $\mathcal{S}$ matrix) for an underlying anyon model in the low energy sector.\label{item:logical}
	\item $R$ is smoothly deformable, that is, the ability to satisfy \ref{item:commuting}-\ref{item:logical} should not depend on the chosen strip of spins defining the support of $R$, given sufficient width.\label{item:deformable}
\end{enumerate}

Since finding a low energy projector appears to be a challenging task, we make the following assumption. Rather than asking for \ref{item:commuting}-\ref{item:deformable} to hold only in the low energy sector, we ask for them to hold on the whole spectrum. At first glance, this seems to be too strong for characterizing TO away from an exactly solvable RG fixed point. Nonetheless, by making this assumption, we can arrive at an effective numerical procedure for detecting TO in 2D models. 
Moreover, in \Sref{S:heuristic}, we provide a heuristic physical justification of this assumption, and some natural avenues for relaxing it in \Sref{S:future}.

\subsection{The Method of Ribbon Operators}\label{S:themethod}

Here we introduce what we call the ribbon operator method for detecting TO. The strategy is simple: we will write a cost function that tries to satisfy the above requirements simultaneously. Since the support of a string-like operator is on a 1D strip, and is expected to be only slightly entangling, we will use the ansatz class of MPOs\cite{Pirvu2010} to describe candidate operators.  We will then use the highly successful methods of 1D systems such as DMRG\cite{Schollwock2011} for optimizing over this ansatz class to find the lowest cost MPO. 

We can then \emph{define} a ribbon operator as any MPO supported on any one-dimensional strip of some fixed width. By drawing on the lattice Hamiltonian and an underlying anyon model, we can define a cost function quantifying the fitness of candidate ribbons given the above criteria. A \emph{good} ribbon operator then corresponds to a local minimum of the cost function.

There is clearly considerable scope within this method for how to use it. The art will be to choose a cost function that is tractable to optimize and gives clear signals of TO when appropriate, such that the distinction between which local minima are ``good'' and ``bad'' is obvious. We do not claim to have a unique or best choice of cost function or optimization routine. The remainder of this section details one particular choice of cost function, and as we show in \Sref{S:Results} this particular choice performs quite well for a variety of simple models.

\subsection{Our Cost Function}\label{S:CostFunction}

We now define our proposed cost function that quantifies how well a candidate ribbon satisfies the above criteria. Let $H=\sum_j h_j$ be some local Hamiltonian and $R$ some candidate ribbon with $\norm{R}=1$ (the Schatten 2-norm, or Frobenius norm $\|R\|^2 = \Tr(R^\dagger R)$) and width $w$, which ideally should be chosen to be comparable to the correlation length. We will also use the symbol $R$ to denote the region of support of the ribbon $R$ since no confusion should result. 

The first condition is that the ribbon should commute with the Hamiltonian of the model. This gives us two terms in the cost function. We can quantify the violation of this condition using $\norm{\comm{R,H}}^2$. This can be decomposed into two distinct contributions by writing the Hamiltonian as
\begin{align}
  H&= H_R + H_{R^{\text{c}}} + H_{\partial R} \nonumber \\
   &= \sum_{j\in R} h_j+\sum_{j\in R^{\text{c}}} h_j+\sum_{j\in \partial R} h_j,
\end{align}
where the first term contains Hamiltonian terms whose support is completely within the support of $R$, the second contains those whose support has no overlap with $R$ and are hence supported entirely in the complementary region $R^{\text{c}}$. The final sum contains all those terms in the Hamiltonian with support on both $R$ and $R^{\text{c}}$. Clearly the second term trivially commutes with $R$ as a consequence of our second condition.

Commutation with the first term cannot be simplified further. However, the term on the boundary $\partial R$ can be simplified as follows. Let $h_j$ be some term on the boundary of the ribbon. Then using the operator Schmidt decomposition, we can split the operator $h_j$ into a sum over interior and exterior components
\begin{align}
  h_j&=\sum_{k} h_{j,k}^{\text{in}}\otimes h_{j,k}^{\text{out}},
\end{align}
where $\Tr\Bigl(h_{j,k}^{\text{out}}{}^\dagger h_{j,k^\prime}^{\text{out}}\Bigr)=\delta_{k,k^\prime}$. 

Using the orthonormality of the $h_{j,k}^{\text{out}}$ terms and the definition of the Frobenius norm, we find that 
\begin{align}
  \norm{\comm{R,h_j}}^2&=\norm{\sum_k\comm{R, h_{j,k}^{\text{in}}}\otimes h_{j,k}^{\text{out}}}^2\nonumber\\
  &=\sum_k\norm{\comm{R, h_{j,k}^{\text{in}}}\otimes h_{j,k}^{\text{out}}}^2\nonumber\\
  &=\sum_k \norm{\comm{R, h_{j,k}^{\text{in}}}}^2,
\end{align}
By appropriate grouping of boundary terms, we can ensure that $\Tr\Bigl(h_{j,k}^{\text{out}}{}^\dagger h_{j^\prime,k^\prime}^{\text{out}}\Bigr)=\delta_{k,k^\prime}\delta_{j,j^\prime}$. (This is equivalent to applying the Schmidt decomposition to $H_{\partial R}$ instead of just one term.) Thus, we can express the boundary contribution as 
\begin{align}
\sum_{j\in\partial R}\sum_k\norm{\comm{R,h_{j,k}^\text{in}}}^2\,. \label{eqn:BoundaryCost}
\end{align}
We find that the commutation condition together with the bounded width condition gives us a term in the cost function proportional to
\begin{align}
  \norm{\comm{R,H}}^2 = \norm{\comm{R,H_R}}^2 + \sum_{j,k\in\partial R}\norm{\comm{R,h_{j,k}^\text{in}}}^2\,.
\end{align}

Because commutation with the Hamiltonian can always be achieved by choosing the ribbon to be the identity operator, we require a competing term to enforce the topological properties such as having a nontrivial $\mathcal{S}$ or $\mathcal{R}$ matrix. For the moment, suppose that we already have a given nontrivial ribbon operator $L$ that crosses $R$ as in \Fref{fig:toricStrings}. (We will discuss how to relax this prior-knowledge assumption below.) We can incorporate anyon data with the following term in the cost function,
\begin{align}
  \bigl\|\comm{R,L}_\eta\bigr\|^2&=\norm{RL-\eta LR}^2\,,\label{eqn:etacomm}
\end{align}
where $\eta$ is a complex number different from 1. 

This choice of topological term in the cost function intuitively reminds us of the $\mathcal{R}$ matrix relations of \Eref{eqn:Rmatrix}. We can see that such a term is in fact sufficient to give a nontrivial $\mathcal{R}$ matrix as follows. Here we specialize to an abelian model, but the discussion could be generalized. Using the $\mathcal{R}$ matrix relation for an abelian model with a fixed total anyon charge $c = a\times b$, we have the relation
\begin{align}
\begin{array}{c}
\includeTikz{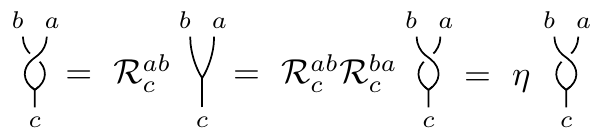}{
    \tikzsetnextfilename{RR}
    \begin{tikzpicture}
    	[a/.style={draw=white, double=black, double distance = .5pt, scale=.6, line width = 1.5pt},
    	b/.style={line width=.5pt, scale=.6,rounded corners = .6em}],
    	\draw[a] {[rounded corners = .75em] (0,.5) -- (.38,.75)} {[rounded corners = .3em] -- (-.2,1.2)} -- (-.2,1.4);
    	\draw[a] {[rounded corners = .75em] (0,.5) -- (-.38,.75)} {[rounded corners = .3em]  -- (.2,1.2)} -- (.2,1.4);
    	\draw[line width=.5, scale=.6] (0,.2) -- (0,.5) -- (-.067,.55) (0,.5) -- (.067,.55);	
    	\node[anchor=north] at (0,.15) {$\scriptstyle c$};
    	\node[anchor=south] at (-.175,.8) {$\scriptstyle b$};
    	\node[anchor=south] at (.175,.8) {$\scriptstyle a$};
    \node[anchor=south] at (.85,.15) {$=\ \mathcal{R}_c^{ab}$};
    \begin{scope}[shift={(1.7,0)}]
    	\draw[b] (0,.7) -- (.2,1)  -- (.2,1.4);
    	\draw[b] (0,.7) -- (-.2,1)  -- (-.2,1.4);
    	\draw[b] (0,.2) -- (0,.7);
    	\node[anchor=north] at (0,.15) {$\scriptstyle c$};
    	\node[anchor=south] at (-.175,.8) {$\scriptstyle b$};
    	\node[anchor=south] at (.175,.8) {$\scriptstyle a$};
    \end{scope}
    \node[anchor=south] at (2.85,.15) {$=\ \mathcal{R}_c^{ab}\mathcal{R}_c^{ba}$};
    \begin{scope}[shift={(4.0,0)}]
            	    	\draw[a] {[rounded corners = .75em] (0,.5) -- (-.38,.75)} {[rounded corners = .3em]  -- (.2,1.2)} -- (.2,1.4);
        	\draw[a] {[rounded corners = .75em] (0,.5) -- (.38,.75)} {[rounded corners = .3em] -- (-.2,1.2)} -- (-.2,1.4);
        	    	\draw[line width=.5, scale=.6] (0,.2) -- (0,.5) -- (-.067,.55) (0,.5) -- (.067,.55);	
        	    	\node[anchor=north] at (0,.15) {$\scriptstyle c$};
        	    	\node[anchor=south] at (-.175,.8) {$\scriptstyle b$};
        	    	\node[anchor=south] at (.175,.8) {$\scriptstyle a$};
        \end{scope}
         \node[anchor=south] at (4.70,.15) {$=\ \eta$};
            \begin{scope}[shift={(5.4,0)}]
                    	    	\draw[a] {[rounded corners = .75em] (0,.5) -- (-.38,.75)} {[rounded corners = .3em]  -- (.2,1.2)} -- (.2,1.4);
                	\draw[a] {[rounded corners = .75em] (0,.5) -- (.38,.75)} {[rounded corners = .3em] -- (-.2,1.2)} -- (-.2,1.4);
                	    	\draw[line width=.5, scale=.6] (0,.2) -- (0,.5) -- (-.067,.55) (0,.5) -- (.067,.55);	
                	    	\node[anchor=north] at (0,.15) {$\scriptstyle c$};
                	    	\node[anchor=south] at (-.175,.8) {$\scriptstyle b$};
                	    	\node[anchor=south] at (.175,.8) {$\scriptstyle a$};
                \end{scope}
    \end{tikzpicture}
}
\end{array}\!\!.
\end{align}
Thus if $\eta \not= 1$ then $\mathcal{R}^{ab}_c$ and $\mathcal{R}^{ba}_c$ cannot both be 1. Note that in the abelian case, $\mathcal{S}_{a\bar{b}}$ can be written as\cite{Kitaev2006} $\mathcal{R}^{ab}\mathcal{R}^{ba}/\mathcal{D}$, so $\eta$ is related to the $\mathcal{S}$ matrix and is therefore gauge invariant (invariant under a change of basis of the fusion space).

Since the anyon charge $c$ is fixed already by the labels $a$ and $b$ in an abelian model, this switching relation should hold even when considering this as just a different operator product order, analogous to the case of a twist product for the $\tilde{\mathcal{S}}$ matrix:
\begin{align}
\begin{array}{c}
\includeTikz{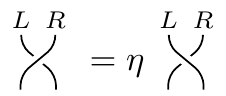}{
    \tikzsetnextfilename{RaRb}
\begin{tikzpicture}
		[a/.style={draw=white, double=black, double distance = .5pt, scale=.6, line width = 1.5pt},
		b/.style={line width=.5pt, scale=.6,rounded corners = .6em}]
	\draw[a] {[rounded corners = .3em] (.3,.5) -- (.3,.75)} {[rounded corners = .3em] -- (-.3,1.25)} -- (-.29,1.4);
	\draw[a] {[rounded corners = .3em] (-.3,.5) -- (-.3,.75)} {[rounded corners = .3em]  -- (.3,1.25)} -- (.29,1.4);	
	\node[anchor=south] at (-.175,.8) {$\scriptstyle L$};
	\node[anchor=south] at (.175,.8) {$\scriptstyle R$};
\node[anchor=south] at (.8,.3) {$=\eta$};
		\begin{scope}[shift={(1.5,0)}]
			\draw[a] {[rounded corners = .3em] (-.3,.5) -- (-.3,.75)} {[rounded corners = .3em]  -- (.3,1.25)} -- (.29,1.4);
				\draw[a] {[rounded corners = .3em] (.3,.5) -- (.3,.75)} {[rounded corners = .3em] -- (-.3,1.25)} -- (-.29,1.4);	
				\node[anchor=south] at (-.175,.8) {$\scriptstyle L$};
				\node[anchor=south] at (.175,.8) {$\scriptstyle R$};
		\end{scope}
\end{tikzpicture}
}
\end{array}\!\!.
\end{align}

Since there are many equivalent strips on which $L$ and $R$ can be supported, the cost should be computed for all possible intersection regions. This is achieved by summing $\bigl\| \comm{R,L}_\eta\bigr\|^2$ over all translates of $L$, labelled $\mathcal{T}(L)$, that lead to an inequivalent intersection with $R$. This translation invariance gives us an additional motivation for our choice of topological term: we can use translation-invariant MPOs in our optimization routines. By comparison, an \emph{a priori} equally attractive term would be to encode the $\mathcal{S}$ matrix relations, but then finite-size effects from the periodic boundary conditions might add additional complications to the numerics. 

Now, we define the $\eta$-\emph{cost} of a ribbon $R$ given some fixed secondary ribbon $L$ as 
\begin{align}
C(R;\eta)=\frac{1}{|R|}
\biggl(\norm{\comm{R,H_R}}^2+&\sum_{j,k\in\partial R}\norm{\comm{R,h^{\text{in}}_{j,k}}}^2
\nonumber\\
&+\sum_{\mathcal{T}(L)}\bigl\|\comm{R,L}_\eta\bigr\|^2\biggr),\label{eqn:CostFunction}
\end{align}
where $|R|$ is the number of spins on which $R$ is supported. Note that due to the nonsymmetric nature of $\comm{\,\cdot\,,\cdot\,}_\eta$, it is convenient to define the $\eta$-cost of $L$ as
\begin{align}
C(L;\eta)=\frac{1}{|L|}\biggl(\norm{\comm{L,H_L}}^2+&\sum_{j,k\in\partial L}\norm{\comm{L,h^{\text{in}}_{j,k}}}^2
\nonumber\\
&+\sum_{\mathcal{T}(R)}\bigl\|\comm{R,L}_\eta\bigr\|^2\biggr),
\end{align}
where the appropriate Hamiltonian terms are taken. Here the normalization term is chosen so that the cost of a ribbon is approximately independent of the volume. 

One could consider summing these two terms to build a total cost function for the pair of ribbons, however we linearize the problem for a fixed $L$ (or fixed $R$) and instead do an alternating minimization algorithm. 

In the case of the exact toric code, the form of the cost function is relatively simple and gives a concrete instantiation of the translation operation $\mathcal{T}$. It is easy to check that the standard logical operators of this code have zero $\eta$-cost when $\eta = -1$, as expected. We present the explicit cost function derivation in Appendix \ref{S:costfunc} for illustration.
%We present the explicit cost function derivation in the Supplementary Material\cite{SM} for illustration.

Although this cost function only involves a pair of ribbons, it is possible to incorporate braid relations with other particle types by adding terms of the form of \Eref{eqn:etacomm} with additional ribbon operators. In this way, it is possible to completely reconstruct the $\mathcal{S}$ matrix of the anyon model up to a column permutation. This ambiguity is due to the difficulty of identifying equivalence of operators on incomparable supports (i.e.\ those with vertical and horizontal supports in \Fref{fig:toricStrings}). We discuss possible avenues to removing this freedom in \Sref{S:future}.

\section{Heuristic Justification for the Norm}\label{S:heuristic}
%------------------------------------------------------------------------------------------------------------%

As discussed in \Sref{S:ribbons}, a key assumption which enables our method to be numerically efficient is that the commutation relations, which are expected to hold only on a topologically ordered ground space, hold on the entire spectrum. While this appears to be a very strong assumption, we believe that our method should continue to work even when this assumption breaks. In this section, we will provide a physically motivated, heuristic justification for this assumption.

At high energies, we expect a gas of short-range interacting anyons. When ribbon operators are used to create additional anyons, braid them, and fuse them in a specific way, the resulting process will in general be affected by the presence of the background anyon gas. Specifically, the expectation value of the braid can be affected by 1) dynamical phases acquired from short range interactions with the background anyons, and 2) topological terms caused by the background anyons enclosed in the braid. To eliminate these effects, we need to make sure that 1) there are no background anyons near the support of the ribbon operator, and 2) the topological charge enclosed in the braid is trivial.

It happens that both of these conditions can be enforced with small width operators. Indeed, 1) requires projecting on the local ground state (LGS), i.e.~the subspace with no charges. Similarly, it is possible to fix the topological charge of a region using an operator which acts only on the boundary of this region and is described by a finite bond dimension MPO.\cite{Haah2014} Let $\Pi_C$ denote the corresponding projector.

Suppose that $R$ is an MPO ribbon operator which obeys the right commutation relations on the ground space, but not on the entire spectrum. Then, $R' = \Pi_C R \Pi_C$ should be a slightly wider and larger-bond-dimension MPO ribbon operator which obeys the commutation on the entire spectrum. Thus, while our approach cannot find $R$, it should be able to find $R'$ without any assumptions.

We stress that the above argument is not key to our method, and no ground state projectors are incorporated directly in our algorithm or in the numerical results presented in \Sref{S:Results}. One could alter the method to explicitly make use of information about the low energy space. We discuss this further in \Sref{S:future}.

\section{Finding Ribbon Operators}\label{S:algorithm}
%------------------------------------------------------------------------------------------------------------%

In many exactly solvable models, an analytic form for the ribbon operators can be found. In more general models, for example ones with a noncommuting Hamiltonian, the ribbons must be found numerically. In this section, we describe how to find ribbon operators using DMRG.\cite{Schollwock2011}

We parameterize a width $w$ ribbon by a block-translationally invariant infinite MPO. This MPO is ``snaked'' along the support of the ribbon to cover the two-dimensional region. That is, the coordinates of the ribbon are treated in a linearized lexicographic order moving along the ribbon lengthwise. This MPO can be vectorized, i.e.\ interpreted as a vector instead of a matrix. An MPO is also built for the cost function, with the same snaking pattern, and standard DMRG can then be applied to find the ribbon that minimizes the constraints.

This vectorization procedure corresponds to the following transformation, written in tensor network diagram notation:
\begin{equation*}
  \includeTikz{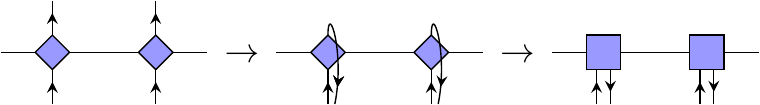}{
  \tikzsetnextfilename{MPOtoMPS}
  \begin{tikzpicture}[a/.style={fill=blue!40},b/.style={fill=blue!40},scale=.7,every node/.style={sloped,allow upside down}]
      \def\a{.5}
      \def\dx{1}
    \foreach \xa in {0,1}{
    \MPOtensor{.5}{1}{\xa};
    }
    \begin{scope}[shift={(3,-.05)}]
    \node at (0,0) {$\to$};
    \end{scope}
    \begin{scope}[shift={(4,0)}]
    \foreach \x in {0,1}{
    \draw[shift={(\x*\a+\x*\dx,0)}] (-\dx/2,0) -- (0,0);
    \filldraw[a,shift={(\x*\a+\x*\dx,0)}] (0,0) -- (\a/2,\a/2) -- (\a,0) -- (\a/2,-\a/2) -- (0,0);=
    \draw[shift={(\x*\a+\x*\dx,0)}] (\a,0) -- (\a+\dx/2,0);
    \draw[shift={(\x*\a+\x*\dx,0)}] (\a/2,\a/2) .. controls (\a/2,\a/2+\dx/2) and (\a,0) ..(\a/2+\a/5,-\a/2-\dx/2);
    \draw[-stealth] (\a/2+\a/5+.05,-.4) -- (\a/2+\a/5+.04,-.5);
    \draw[shift={(\x*\a+\x*\dx,0)},-stealth] (\a/2+\a/5+.05,-.4) -- (\a/2+\a/5+.04,-.5);
    \draw[shift={(\x*\a+\x*\dx,0)}] (\a/2,-\a/2-\dx/2) -- node {\midarrow} (\a/2,-\a/2);
    }
    \end{scope}
     \begin{scope}[shift={(7,-.05)}]
        \node at (0,0) {$\to$};
        \end{scope}
  \begin{scope}[shift={(8,-.25)}]
    \foreach \x in {0,1}{
    \draw[shift={(\x*\a+\x*\dx,0)}] (-\dx/2,\a/2) -- (0,\a/2);
    \filldraw[b,shift={(\x*\a+\x*\dx,0)}] (0,0) -- (0,\a) -- (\a,\a) -- (\a,0) -- (0,0);
    \draw[shift={(\x*\a+\x*\dx,0)}] (\a,\a/2) -- (\a+\dx/2,\a/2);
    \draw[shift={(\x*\a+\x*\dx,0)}] (\a/2-\a/5,-\dx/2) -- node {\midarrow}(\a/2-\a/5,0);
    \draw[shift={(\x*\a+\x*\dx,0)}] (\a/2+\a/5,0) -- node {\midarrow} (\a/2+\a/5,-\dx/2);
    }
    \end{scope}
  \end{tikzpicture}}.
\end{equation*}
Given any operator $O$ in MPO form, we can explicitly write the form of the constraint term $\norm{\comm{R,O}}^2$ for one site of the MPO as
\begin{align}
  \begin{array}{l}
  \includeTikz{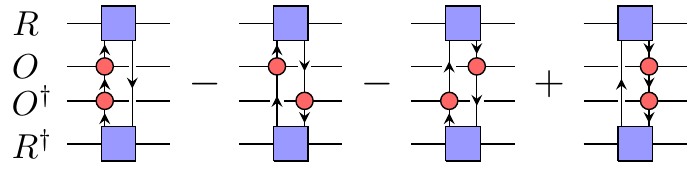}{
    \tikzsetnextfilename{CostFunctionMPSForm}
    \begin{tikzpicture}[a/.style={fill=blue!40},
      b/.style={fill=red!60},
      c/.style={fill=red!10},
      scale=.7,
      every node/.style={sloped,allow upside down}]
      \def\a{.5}
      \def\dx{1}
      \def\dy{1}
      \def\a{.5}
      \def\dx{1}
      \path[shift={(-\dx-.1,\a/2)}] (0,0) node{$R$};
      \path[shift={(-\dx-.1,-\dx/4-\a/4)}] (0,0) node{$O$};
      \path[shift={(-\dx,-\dx/4-5*\a/4)}] (0,0) node{$O^\dagger$};
      \path[shift={(-\dx,-\dx/4-5*\a/2)}] (0,0) node{$R^\dagger$};
      \draw[shift={(0,0)}] (-\dx/2,\a/2) -- (0,\a/2);
      \filldraw[a,shift={(0,0)}] (0,0) -- (0,\a) -- (\a,\a) -- (\a,0) -- (0,0);
      \draw[shift={(0,0)}] (\a,\a/2) -- (\a+\dx/2,\a/2);
      \draw[shift={(0,0)}] (9*\a/10,0)-- node {\midarrow}(9*\a/10,-\dx/4-2*\a);
      \draw[shift={(0,0)}] (\a/10,-\dx/4) -- node {\midarrow} (\a/10,0);
      \filldraw[b,shift={(0,0)}] (\a/10,-\dx/4-\a/4) circle (\a/4);
      \draw[shift={(0,0)}] (\a/10+\a/4,-\dx/4-\a/4)--(8*\a/10,-\dx/4-\a/4);
      \draw[shift={(0,0)}] (\a,-\dx/4-\a/4)--(\a+\dx/2,-\dx/4-\a/4);
      \draw[shift={(0,0)}] (-\dx/2,-\dx/4-\a/4)--(\a/10-\a/4,-\dx/4-\a/4);
      \draw[shift={(0,0)}] (\a/10,-\dx/4-\a) -- node {\midarrow} (\a/10,-\dx/4-\a/2);
      \filldraw[b,shift={(0,0)}] (\a/10,-\dx/4-5*\a/4) circle (\a/4);
      \draw[shift={(0,0)}] (\a/10+\a/4,-\dx/4-5*\a/4)--(8*\a/10,-\dx/4-5*\a/4);
      \draw[shift={(0,0)}] (\a,-\dx/4-5*\a/4)--(\a+\dx/2,-\dx/4-5*\a/4);
      \draw[shift={(0,0)}] (-\dx/2,-\dx/4-5*\a/4)--(\a/10-\a/4,-\dx/4-5*\a/4);
      \draw[shift={(0,0)}] (\a/10,-\dx/4-8*\a/4) -- node {\midarrow} (\a/10,-\dx/4-6*\a/4);
      \filldraw[a,shift={(0,-\dx/4-3*\a)}] (0,0) -- (0,\a) -- (\a,\a) -- (\a,0) -- (0,0);
      \draw[shift={(0,-\dx/4-3*\a)}] (-\dx/2,\a/2) -- (0,\a/2);
      \draw[shift={(0,-\dx/4-3*\a)}] (\a,\a/2) -- (\a+\dx/2,\a/2);
      %----------------------------------------------------------------------------
      \path (\a+\dx,-\dx/8-\a) node{\large $-$};
      \draw[shift={(\a+2*\dx,0)}] (-\dx/2,\a/2) -- (0,\a/2);
      \filldraw[a,shift={(\a+2*\dx,0)}] (0,0) -- (0,\a) -- (\a,\a) -- (\a,0) -- (0,0);
      \draw[shift={(\a+2*\dx,0)}] (\a,\a/2) -- (\a+\dx/2,\a/2);
      \draw[shift={(\a+2*\dx,0)}] (9*\a/10,0)-- node {\midarrow}(9*\a/10,-\dx/4-\a);
      \draw[shift={(\a+2*\dx,0)}] (9*\a/10,-\dx/4-6*\a/4)-- node {\midarrow}(9*\a/10,-\dx/4-2*\a);
      \draw[shift={(\a+2*\dx,0)}] (\a/10,-\dx/4) -- node {\midarrow} (\a/10,0);
      \filldraw[b,shift={(\a+2*\dx,0)}] (\a/10,-\dx/4-\a/4) circle (\a/4);
      \draw[shift={(\a+2*\dx,0)}] (\a/10+\a/4,-\dx/4-\a/4)--(7*\a/10,-\dx/4-\a/4);
      \draw[shift={(\a+2*\dx,0)}] (11*\a/10,-\dx/4-\a/4)--(\a+\dx/2,-\dx/4-\a/4);
      \draw[shift={(\a+2*\dx,0)}] (-\dx/2,-\dx/4-\a/4)--(\a/10-\a/4,-\dx/4-\a/4);
      \filldraw[b,shift={(\a+2*\dx,0)}] (9*\a/10,-\dx/4-5*\a/4) circle (\a/4);
      \draw[shift={(\a+2*\dx,0)}] (9*\a/10-\a/4,-\dx/4-5*\a/4)--(3*\a/10,-\dx/4-5*\a/4);
      \draw[shift={(\a+2*\dx,0)}] (9*\a/10+\a/4,-\dx/4-5*\a/4)--(\a+\dx/2,-\dx/4-5*\a/4);
      \draw[shift={(\a+2*\dx,0)}] (-\a/10,-\dx/4-5*\a/4)--(-\dx/2,-\dx/4-5*\a/4);
      \draw[shift={(\a+2*\dx,0)}] (\a/10,-\dx/4-8*\a/4) -- node {\midarrow} (\a/10,-\dx/4-\a/2);
      \filldraw[a,shift={(\a+2*\dx,-\dx/4-3*\a)}] (0,0) -- (0,\a) -- (\a,\a) -- (\a,0) -- (0,0);
      \draw[shift={(\a+2*\dx,-\dx/4-3*\a)}] (-\dx/2,\a/2) -- (0,\a/2);
      \draw[shift={(\a+2*\dx,-\dx/4-3*\a)}] (\a,\a/2) -- (\a+\dx/2,\a/2);
      %----------------------------------------------------------------------------
      \path (2*\a+3*\dx,-\dx/8-\a) node{\large $-$};
      \draw[shift={(2*\a+4*\dx,0)}] (-\dx/2,\a/2) -- (0,\a/2);
      \filldraw[a,shift={(2*\a+4*\dx,0)}] (0,0) -- (0,\a) -- (\a,\a) -- (\a,0) -- (0,0);
      \draw[shift={(2*\a+4*\dx,0)}] (\a,\a/2) -- (\a+\dx/2,\a/2);
      \draw[shift={(2*\a+4*\dx,0)}] (\a/10,-\dx/4-\a)-- node {\midarrow}(\a/10,0);
      \draw[shift={(2*\a+4*\dx,0)}] (\a/10,-\dx/4-2*\a)-- node {\midarrow}(\a/10,-\dx/4-6*\a/4);
      \draw[shift={(2*\a+4*\dx,0)}] (9*\a/10,0) -- node {\midarrow} (9*\a/10,-\dx/4);
      \draw[shift={(2*\a+4*\dx,0)}] (9*\a/10,-\dx/4-\a/2) -- node {\midarrow} (9*\a/10,-\dx/4-8*\a/4);
      \filldraw[b,shift={(2*\a+4*\dx,0)}] (9*\a/10,-\dx/4-\a/4) circle (\a/4);
      \draw[shift={(2*\a+4*\dx,0)}] (7*\a/10,-\dx/4-5*\a/4)--(\a/10+\a/4,-\dx/4-5*\a/4);
      \draw[shift={(2*\a+4*\dx,0)}] (11*\a/10,-\dx/4-5*\a/4)--(\a+\dx/2,-\dx/4-5*\a/4);
      \draw[shift={(2*\a+4*\dx,0)}] (\a/10-\a/4,-\dx/4-5*\a/4)--(-\dx/2,-\dx/4-5*\a/4);
      \filldraw[b,shift={(2*\a+4*\dx,0)}] (\a/10,-\dx/4-5*\a/4) circle (\a/4);
      \draw[shift={(2*\a+4*\dx,0)}] (9*\a/10-\a/4,-\dx/4-\a/4)--(3*\a/10,-\dx/4-\a/4);
      \draw[shift={(2*\a+4*\dx,0)}] (9*\a/10+\a/4,-\dx/4-\a/4)--(\a+\dx/2,-\dx/4-\a/4);
      \draw[shift={(2*\a+4*\dx,0)}] (-\dx/2,-\dx/4-\a/4)--(-\a/10,-\dx/4-\a/4);
      \filldraw[a,shift={(2*\a+4*\dx,-\dx/4-3*\a)}] (0,0) -- (0,\a) -- (\a,\a) -- (\a,0) -- (0,0);
      \draw[shift={(2*\a+4*\dx,-\dx/4-3*\a)}] (-\dx/2,\a/2) -- (0,\a/2);
      \draw[shift={(2*\a+4*\dx,-\dx/4-3*\a)}] (\a,\a/2) -- (\a+\dx/2,\a/2);
      %---------------------------------------------------------------------------------------
      \path (3*\a+5*\dx,-\dx/8-\a) node{\large $+$};
      \draw[shift={(3*\a+6*\dx,0)}] (-\dx/2,\a/2) -- (0,\a/2);
      \filldraw[a,shift={(3*\a+6*\dx,0)}] (0,0) -- (0,\a) -- (\a,\a) -- (\a,0) -- (0,0);
      \draw[shift={(3*\a+6*\dx,0)}] (\a,\a/2) -- (\a+\dx/2,\a/2);
      \draw[shift={(3*\a+6*\dx,0)}] (\a/10,-\dx/4-2*\a)-- node {\midarrow}(\a/10,0);
      \draw[shift={(3*\a+6*\dx,0)}] (9*\a/10,0) -- node {\midarrow} (9*\a/10,-\dx/4);
      \filldraw[b,shift={(3*\a+6*\dx,0)}] (9*\a/10,-\dx/4-\a/4) circle (\a/4);
      \draw[shift={(3*\a+6*\dx,0)}] (9*\a/10+\a/4,-\dx/4-\a/4)--(\a+\dx/2,-\dx/4-\a/4);
      \draw[shift={(3*\a+6*\dx,0)}] (9*\a/10-\a/4,-\dx/4-\a/4)--(2*\a/10,-\dx/4-\a/4);
      \draw[shift={(3*\a+6*\dx,0)}] (0,-\dx/4-\a/4)--(-\dx/2,-\dx/4-\a/4);
      \draw[shift={(3*\a+6*\dx,0)}] (9*\a/10,-\dx/4-\a/2) -- node {\midarrow} (9*\a/10,-\dx/4-\a);
      \filldraw[b,shift={(3*\a+6*\dx,0)}] (9*\a/10,-\dx/4-5*\a/4) circle (\a/4);
      \draw[shift={(3*\a+6*\dx,0)}] (9*\a/10+\a/4,-\dx/4-5*\a/4)--(\a+\dx/2,-\dx/4-5*\a/4);
      \draw[shift={(3*\a+6*\dx,0)}] (9*\a/10-\a/4,-\dx/4-5*\a/4)--(2*\a/10,-\dx/4-5*\a/4);
      \draw[shift={(3*\a+6*\dx,0)}] (0,-\dx/4-5*\a/4)--(-\dx/2,-\dx/4-5*\a/4);
      \draw[shift={(3*\a+6*\dx,0)}] (9*\a/10,-\dx/4-6*\a/4) -- node {\midarrow} (9*\a/10,-\dx/4-8*\a/4);
      \filldraw[a,shift={(3*\a+6*\dx,-\dx/4-3*\a)}] (0,0) -- (0,\a) -- (\a,\a) -- (\a,0) -- (0,0);
      \draw[shift={(3*\a+6*\dx,-\dx/4-3*\a)}] (-\dx/2,\a/2) -- (0,\a/2);
      \draw[shift={(3*\a+6*\dx,-\dx/4-3*\a)}] (\a,\a/2) -- (\a+\dx/2,\a/2);
    \end{tikzpicture}}
    \end{array}.
  \end{align}
By concatenating this expression and closing the boundaries appropriately, we see that this is equivalent to the expectation of some MPO for some matrix product ``state'' given by the vectorization of $R$.

The total cost function can be obtained by summing up the MPOs defining each cost term, corresponding to direct sum on the MPO matrices. The cost function MPO will not generally correspond to a local Hamiltonian, which is the regime where DMRG is usually applied. Indeed, due to the squaring required on the internal commutator, we will have terms like $h_j h_k$ for every pair $j,\,k$. Despite this potential challenge, we will demonstrate that DMRG is an effective algorithm for this minimization.
  
\subsection{Optimizing}

We attempt to optimize $C(R;\eta)$ for various choices of $\eta$. In particular, we restrict ourselves to the unit disk as this is sufficient for the models below, although extending to other $\eta$ does not change any of the results obtained. 

When attempting to optimize a pair, we used alternating minimization. First $C(R_1;\eta)$ was minimized in the presence of a random $R_2$. Then $C(R_2;\eta)$ was optimized given the $R_1$ obtained from the previous optimization. The $R_2$ obtained was then used to seed the next iteration. This process was repeated until convergence.

For each choice of $\eta$, several (usually 5 or 10) ribbons or pairs of ribbons were optimized from random initial pairs and the lowest cost pair was retained. This proved necessary to avoid premature convergence to a local minimum, particularly in the alternating minimization, leading to a pair that was orthogonal to the optimal solution. In most cases a total of 5-10 restarts appeared sufficient to ensure that an optimal pair was obtained. For each value of $\eta$, an independent random start was generated, rather than using a warm start from the previous nearby values of $\eta$. 

\section{Numerical Results}\label{S:Results}
%------------------------------------------------------------------------------------------------------------%

In this section we use ribbon operators to study several standard models that exhibit both topologically ordered and trivial phases. We demonstrate the efficacy of ribbon operators in identifying TO both at exactly solvable points and away from such points. 

\subsection{\texorpdfstring{$\ZZ_d$}{Z_d} Quantum Double-Ising model}\label{S:QDIResults}

\begin{figure}
\centering
\includeTikz{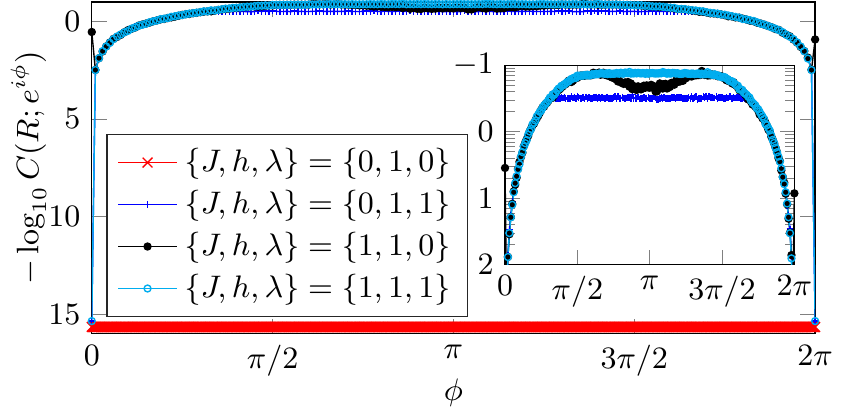}{
  \tikzsetnextfilename{TopologicallyTrivial}
  \setlength\figureheight{.39\columnwidth} 
  \setlength\figurewidth{.85\columnwidth} 
  \input{graphs/TopologicallyTrivial.tikz}}
  \caption{Cost for ribbons in topologically trivial phases the QDI$_2$ model with $\eta=\mathrm{e}^{i\phi}$. A pair of ribbons were optimized on two $w=1$ strips with bond dimension $1$ for the paramagnet and $2$ in the other cases. We observe two distinct failures to observe topological order. In the case of the paramagnet, a zero-cost ribbon can be found for all $\phi$. In fact, this can be extended to all $\eta$. In the other cases (closeup in inset plot), there is no $\eta$ for which a nontrivial low-cost ribbon can be obtained. It is possible that this could be due to insufficient width, although we have not observed this to be the case. Note that averaging over many trials (20) to give a typical picture of the algorithms performance. In the case where the variational minimum is used instead, still no signal of topological order is observed.}\label{fig:TopologicallyTrivialCostData}
  \vspace*{-3mm}
\end{figure}

We define the $\ZZ_d$ Quantum Double-Ising (QDI$_d$) model as follows.\cite{Kitaev2003, Brown2014} Let $\Lambda$ be the  bicolorable lattice shown in \Fref{fig:toricStrings}. Place a qudit ($d$ level quantum system) at each vertex and define the generalized unitary Pauli operators $X$ and $Z$ such that $ZX=\omega XZ$, $X^d = Z^d =\mathbbm{1}$, and $\omega=\mathrm{e}^{2\pi i/d}$. We define plaquette operators $a$ and $b$ as 
\begin{align}
\begin{array}{ccc}
\includeTikz{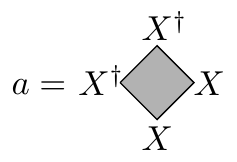}{
\tikzsetnextfilename{ZnAOperator}
\begin{tikzpicture}    [a/.style={fill=black!30},
        e/.style={fill=black!5,opacity=1},
        scale=.75]
    \def\dx{1}
    \def\dy{1};
    \filldraw[a] (-\dx/2,0) {} --(0,\dy/2)--(\dx/2,0)--(0,-\dy/2)--(-\dx/2,0);
\node[anchor=base] at (-\dx-.6,-.165) {$a=$};
    \begin{scope}[shift={(-.025,0)}]
    \node[anchor=south west] at (-.35,.4) {$X^\dagger$};
     \node[anchor=south west,shift={(.38,-.38)}] at (-.15,.17) {$X$};
    \node[anchor=south west,shift={(0,-.75)}] at (-.35,-.08) {$X$};
    \node[anchor=south west,shift={(-.38,-.38)}] at (-.7,.17) {$X^\dagger$};
    \end{scope}
  \end{tikzpicture}}
  &&
  \includeTikz{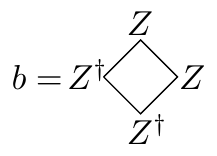}{
  \tikzsetnextfilename{ZnBOperator}
  \begin{tikzpicture}[a/.style={fill=black!0},
          e/.style={fill=black!5,opacity=1},
          scale=.75]
    \def\dx{1}
    \def\dy{1};
    \filldraw[a] (-\dx/2,0) {} --(0,\dy/2)--(\dx/2,0)--(0,-\dy/2)--(-\dx/2,0);
\node[anchor=base] at (-\dx-.4,-.175) {$b=$};
	\node[anchor=south west] at (-.35,.4) {$Z$};
    \node[anchor=south west,shift={(.38,-.38)}] at (-.15,.17) {$Z$};
    \node[anchor=south west,shift={(0,-.75)}] at (-.35,-.08) {$Z^\dagger$};
    \node[anchor=south west,shift={(-.38,-.38)}] at (-.65,.17) {$Z^\dagger$};
  \end{tikzpicture}}
  \end{array},
\end{align}
on the dark and light plaquettes respectively. The Hamiltonian is given by
\begin{subequations}\label{eqn:QDI}
\begin{alignat}{2}
H_{\ZZ_d}=&-J\sum_{p\in\{\raisebox{-1.7pt}{\includeTikz{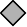}{\tikzsetnextfilename{DarkPlaquette}\tikz[scale=.25,baseline=-2pt] \filldraw[fill=black!30](-.5,0)--(0,.5)--(.5,0)--(0,-.5)--(-.5,0);}}\hspace*{-1.3mm}\}}P_p-J\sum_{p\in \{\raisebox{-1.7pt}{\includeTikz{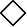}{\tikzsetnextfilename{LightPlaquette}\tikz[scale=.25,baseline=-2pt] \filldraw[fill=black!00](-.5,0)--(0,.5)--(.5,0)--(0,-.5)--(-.5,0);}}\hspace*{-1.3mm}\}}Q_p \label{subeq:TO}\\
 & -\frac{h}{2}\sum_j (X_j+X_j^\dagger) \label{subeq:PM}\\
& -\frac{\lambda}{4}\sum_{\langle j,k\rangle} (Z_j+Z_j^\dagger)(Z_k+Z_k^\dagger),\label{subeq:FM}
\end{alignat}
\end{subequations}
where at each plaquette $p$ we have the operators
\begin{align}
 P=\sum_{k=1}^{d-1} a^k\ , \quad Q=\sum_{k=1}^{d-1} b^k \,.
\end{align}
The Hamiltonian contains the topological terms in \Eref{subeq:TO} with strength $J$, a transverse onsite $X$ field in \Eref{subeq:PM} with strength $h$, and a ferromagnetic Ising-type term in \Eref{subeq:FM} with strength $\lambda$.

\begin{figure}
\centering
  \includeTikz{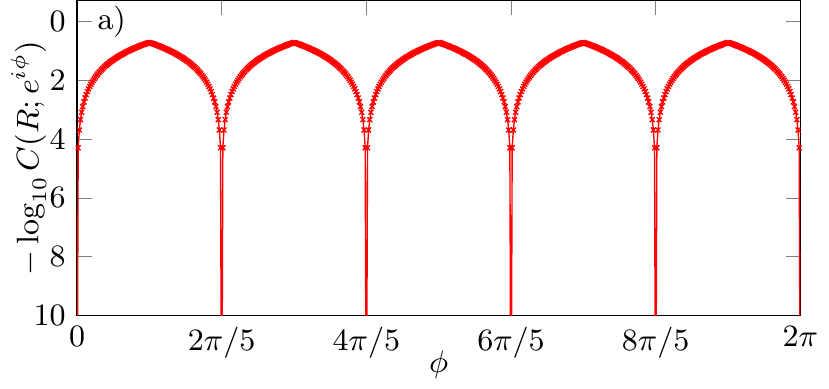}{
  \tikzsetnextfilename{Z5QD}
  \setlength\figureheight{.37\columnwidth} 
  \setlength\figurewidth{.85\columnwidth}
  \input{graphs/Z5.tikz}}\\
  \includeTikz{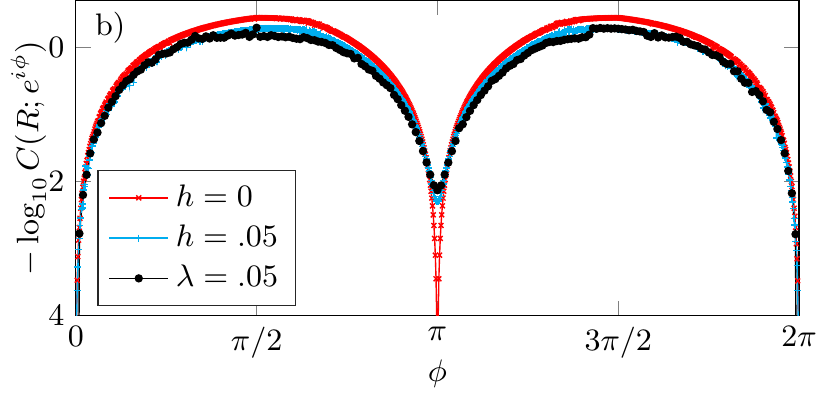}{
  \tikzsetnextfilename{Z2QD}
  \setlength\figureheight{.37\columnwidth} 
  \setlength\figurewidth{.85\columnwidth} 
  \input{graphs/Z2.tikz}}\\
  \includeTikz{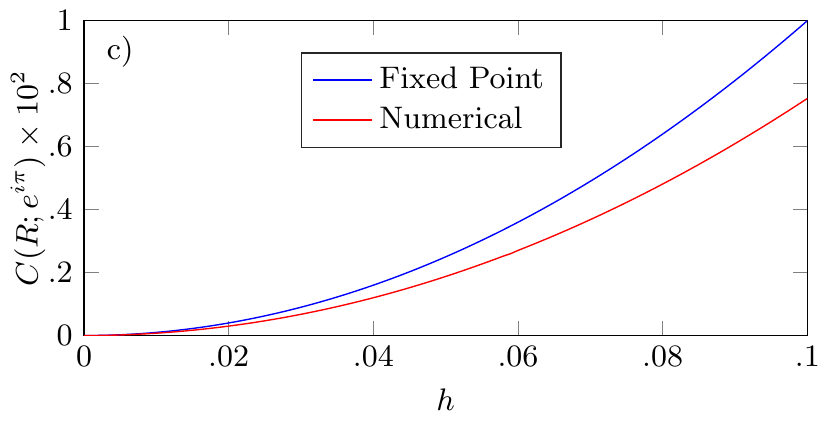}{
  \tikzsetnextfilename{comptofp}
  \setlength\figureheight{.37\columnwidth} 
  \setlength\figurewidth{.85\columnwidth} 
  \input{CompareToFP.tikz}}
  \caption{a) Cost for a width 1 ribbon found in the QDI$_5$ model of \Eref{eqn:QDI} with $\{J,h,\lambda\}=\{1,0,0\}$ using bond dimension 5 and plotted with $\eta=\mathrm{e}^{i\phi}$. The other ribbon was kept as a string of $Z$s as shown in \Fref{fig:toricStrings}.\\
  b) Cost for ribbons found using alternating minimization on the QDI$_2$ model with $J=1$. In the $h=0$ case a pair of $w=1$, bond dimension 1 ribbons were sought, in the other cases, the pair consisted of a $w=1$ and a $w=2$ ribbon with bond dimensions $1$ and $5$ respectively. At the exactly solvable points ($h=0$), we see a dramatic cost decrease at $d$th roots of unity, signalling both the presence of and type of topological order present in these models. This feature remains even with the addition of an $X$ field or Ising type term which destroys the solvability of the model.\\
  c) Cost of a width 4 ribbon of bond dimension 2 in the QDI$_2$ model. The cost obtained from our numerical optimization is smaller than that of the fixed point string for all (nonzero) values of $h$. This demonstrates the nontrivial nature of the ribbon operators we obtain, even in this perturbative regime.}\label{fig:ZQDICostData}
  \vspace*{-5mm}
\end{figure}

When $h=\lambda=0$, this is a topologically ordered commuting model which reduces to the toric code when $d=2$. This model also captures a number of other models, including the paramagnet ($\{J,h,\lambda\}=\{0,1,0\}$) and a ferromagnet ($\{J,h,\lambda\}=\{0,0,1\}$). These latter correspond to distinct topologically trivial phases. At the fixed point of each of these phases, the model is exactly solvable. In the $d=2$ case, the generic model with $J=0$ is given by the Hamiltonian
\begin{align}
  H=-h\sum_j X_j-\lambda\sum_{\langle j,k\rangle} Z_jZ_k\,.
\end{align}
At the RG fixed points of these phases (corresponding to either $h=0$ or $\lambda=0$), we can set $R_1=\prod \ket{\uparrow}\bra{\uparrow}$ and $R_2=\prod \ket{\downarrow}\bra{\downarrow}$, where $\ket{\uparrow}$/$\ket{\downarrow}$ correspond to the $+1$/$-1$ eigenstates of the relevant operator (i.e.\ the eigenstates of $X$ when $\lambda=0$ and of $Z$ when $h=0$). These ribbons commute with the Hamiltonian. Since $R_1R_2=R_2R_1=0$, their $\eta$-commutation relation is not unique, so they do not correspond to good ribbon operators. We therefore expect to be able to find zero-cost ribbons for all $\eta$ at these points. We note that constraining the ribbons to be unitary would be a natural way to eliminate these spurious solutions in the trivial phases, but this is computationally expensive; we discuss this point further in \Sref{S:future}. 

Costs for ribbons at various non-topological points in the QDI$_d$ model are shown in \Fref{fig:TopologicallyTrivialCostData}. We observe two distinct ways in which trivial order can be signalled. As we have already discussed, one of these corresponds to mutually annihilating operators as in the para- and ferromagnetic phases.  In the other case we simply fail to observe any low cost ribbons which form a nontrivial algebra. There are two explanations for this. Either there are no ribbon operators to be found or the chosen width $w$ is too narrow. 

\begin{figure}
\includeTikz{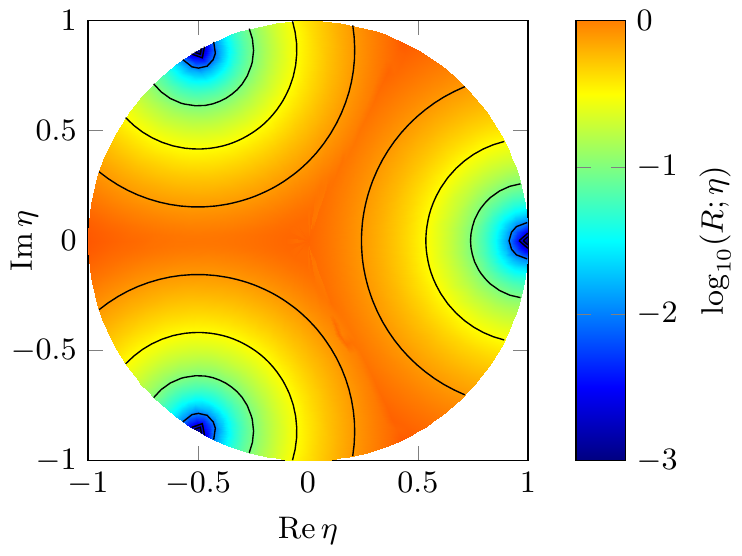}{
  \setlength\figureheight{.7\columnwidth} 
  \setlength\figurewidth{.7\columnwidth} 
  \tikzsetnextfilename{Z3Disk} 
  \input{graphs/Z3Disk.tikz}}
 \caption{Cost for ribbons found using alternating minimization on the QDI$_3$ model with $\{J,h,\lambda\}=\{1,0,0\}$. A pair of  $w=1$ ribbons with bond dimension 1 were optimized. We observe a signal of topological order only at $\eta=\mathrm{e}^{2\pi i k/3}$.}\label{fig:Z3DiskCostData}
 \vspace*{-5mm}
\end{figure}
\begin{figure}[h!]
\centering
\includeTikz{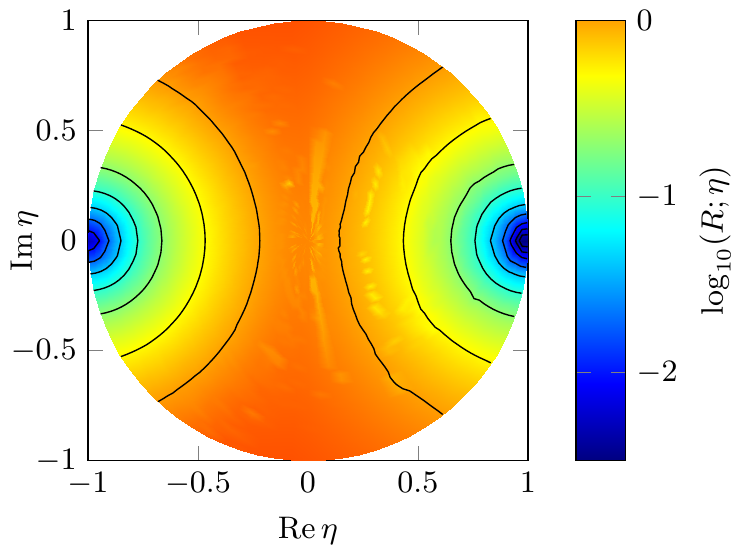}{
  \setlength\figureheight{.7\columnwidth} 
  \setlength\figurewidth{.7\columnwidth} 
  \tikzsetnextfilename{Z2Disk} 
  \input{graphs/TCDisk.tikz}}
  \caption{Cost for ribbons found using alternating minimization on the QDI$_2$ model with $\{J,h,\lambda\}=\{1,0.05,0\}$ where $\eta$ runs over points in the complex unit disk. A pair of ribbons with $w=1$ and $w=2$ and bond dimensions 1 and 5 respectively were optimized and the total cost is shown. We observe a signal of topological order at $\eta=-1$. Note that \Fref{fig:ZQDICostData}b corresponds to the edge of this disk. The raw data is displayed, but some contours were subjected to gaussian smoothing for clarity in the nearly flat region where the noise in the data and the gradient of the function are comparable in magnitude.}\label{fig:Z2DiskCostData}
  \vspace*{-5mm}
\end{figure}

The phase which includes the point $\{J,h,\lambda\}=\{1,0,0\}$ is topologically ordered, and we expect it to show a signal of $\ZZ_d$-topological order. We therefore expect to observe low cost ribbons only when $\eta=\mathrm{e}^{i\phi}$ is a $d$th root of 1. This property is expected to persist away from the exactly solvable point. In \Fref{fig:ZQDICostData} we show the data obtained at the fixed points of the $\ZZ_5$ and $\ZZ_2$ models and when an $X$ field is turned on in the latter. As expected, at the fixed point a very strong signal is observed. Using a large density of points we see that the cost drops dramatically over a very narrow region around the $d$th root of unity. When we turn on the field term, the signal decreases in magnitude but is still unmistakable, with more than two orders of magnitude in total contrast.
Despite the clear signal, we observe that the value of the cost function at the minimum is comparable to the (square of) the perturbation strength. This may lead one to suspect that the method is simply returning the analytically obtainable fixed point ribbon. To refute this hypothesis, in Fig.~\hyperref[fig:ZQDICostData]{\ref*{fig:ZQDICostData}c}, we compare the cost of numerically obtained ribbons to those of the fixed point strings. Encouragingly, we see that the cost is smaller for the optimized ribbon, indicating that the method is finding a nontrivial result.

In \Fref{fig:Z3DiskCostData}, we present cost data for the fixed point of the QDI$_3$ model over the entire unit disk. As expected, low cost ribbons can only be found around the third root of unity, with no minima occurring inside the disk. This property persists even when an $X$ field is turned on as shown in \Fref{fig:Z2DiskCostData}. This figure shows the cost of ribbons in the QDI$_2$ model with $\{J,h,\lambda\}=\{1,0.05,0\}$. Even at this nonintegrable point, no low cost ribbons can be found anywhere away from the $d$th roots of unity.

\subsection{The Quantum Compass Model}
It is interesting to consider the 2D quantum compass (QC) or Bacon-Shor model\cite{Dorier2005, Bacon2006} in the context of ribbon operators. This model is defined on a square lattice $\Lambda$ with Hamiltonian
\begin{align}\label{eqn:QCmodel}
H_{\mathrm{QC}}&=-J\sum_{i,j\in \Lambda}\left(X_{i,j}X_{i,j+1}+Z_{i,j}Z_{i+1,j}\right).
\end{align}

This model fails to exhibit many of the features expected of a topologically ordered model. For example, the ground state degeneracy does not depend on the underlying lattice topology. Moreover, this appears to be a gapless model with exponential ground state degeneracy.\cite{Dorier2005} Thus, this Hamiltonian is usually described as not having topological order. However, this model can be used to encode a qubit which is protected against arbitrary local errors, having a code distance extensive in the (linear) lattice size,\cite{Bacon2006} a property which is clearly required of a topological code. This qubit can be defined by the pair of logical operators $\tilde{Z}_i=\prod_{j} Z_{i,j}$ and $\tilde{X}_j=\prod_{i} X_{i,j}$. These operators commute with each Hamiltonian term, and intersect at a single point, hence they anticommute, $[\tilde{Z}_i,\tilde{X}_j]_{-1}=0$, and therefore almost fulfill the conditions we give to be good ribbon operators. We will see below that they are rigid objects that exist due to special lattice symmetries, and they are not deformable. 

The Hamiltonian \Eref{eqn:QCmodel} also commutes with all elements of the \emph{stabilizer group} 
\begin{align}
\mathcal{G}&=\left<\prod_j Z_{i,j}Z_{i+1,j},\prod_j X_{j,k}X_{j,k+1}\right>.
\end{align}
It is therefore clear that $\tilde{Z}_j$($\tilde{X}_j$) is related to $\tilde{Z}_k$($\tilde{X}_k$) by application of a stabilizer element, and so defines the same encoded qubit in the ground space. Hence, the logical operators can be moved along to parallel strips. 

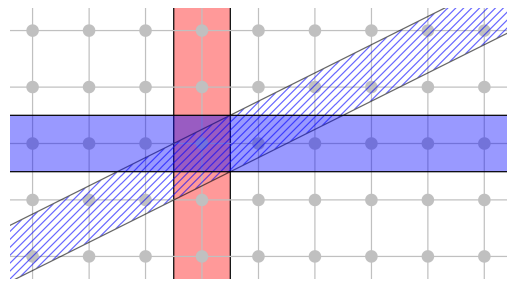
\begin{figure}
  \centering
  \includeTikz{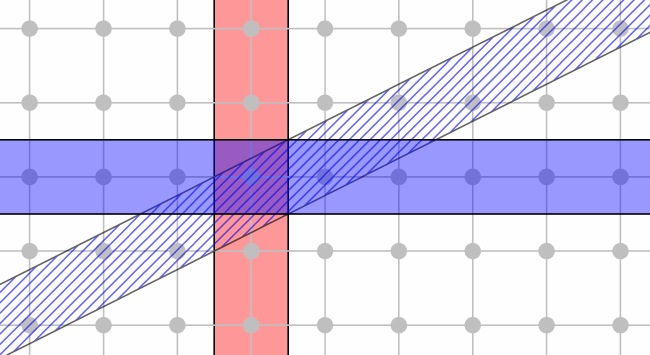}{
  \tikzsetnextfilename{RotatedQCM}
  \begin{tikzpicture}
   [ b/.style={fill=black!20},
          c/.style={fill=red,fill opacity = .4},
          d/.style={fill=blue,fill opacity = .4},
          e/.style={pattern color=blue,opacity=.6,pattern=north east lines},
          f/.style={fill=black!30,draw=black!30},
          scale=.75]
    \def\maxx{10};
    \def\maxy{10};
    \def\a{1};
    \def\r{.1};
    \clip (.6,.6) rectangle (9.4,5.4);
    \filldraw[c] (3.5,.5)--(4.5,.5)--(4.5,6.5)--(3.5,6.5)--(3.5,.5);
    \foreach \x in {0,1,...,\maxx}{
    \foreach \y in {0,1,...,\maxy}{
	   \draw[f,shift={(\x,\y)}] (0,0)--(\a,0);
	   \draw[f,shift={(\x,\y)}] (0,0)--(0,\a);
	   \fill[f,shift={(\x,\y)}] (0,0) circle (\r);
	}
	}
    \filldraw[e] (.5,.5)--(10.5,5.5)--(10.5,6.5)--(.5,1.5)--(.5,.5);
    \filldraw[d] (.1,2.5)--(10.5,2.5)--(10.5,3.5)--(.1,3.5)--(.1,2.5);
  \end{tikzpicture}}
  \caption{Topologically ordered models should support low-cost ribbons on both the horizontal and the rotated blue regions (or some widening of them). This corresponds to insensitivity of the anyon braid relations to the specific worldlines they trace out. Models with specific lattice symmetries such as the quantum compass model might provide false signatures of topological order if ribbon operators are sought only on non-generic strips of the lattice.}\label{fig:RotatedRibbons}
\end{figure}

To investigate topological order in the QC model, we seek to find logical operators supported on rotated regions as shown in \Fref{fig:RotatedRibbons} or widened versions thereof. We compare this to the $\ZZ_2$ quantum double model where topological order is well understood. 

\begin{figure}
\centering
  \includeTikz{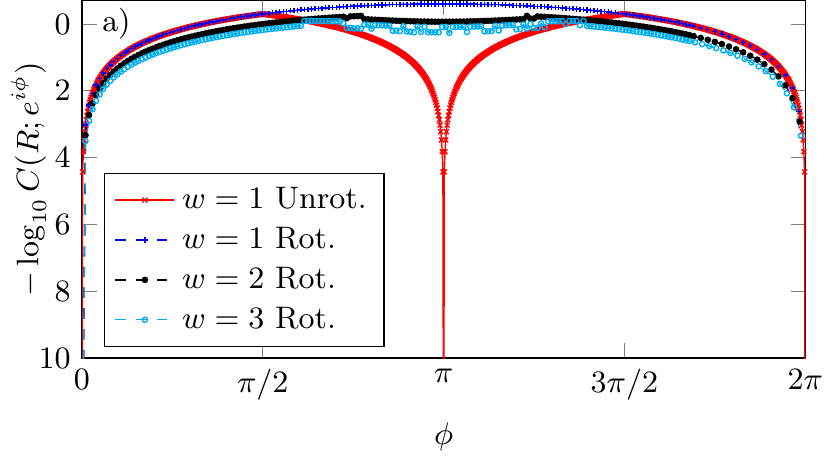}{
  \tikzsetnextfilename{RotatedQCMCostData}
  \setlength\figureheight{.42\columnwidth} 
  \setlength\figurewidth{.85\columnwidth}
  \input{graphs/QuantumCompass.tikz}}\\
  \includeTikz{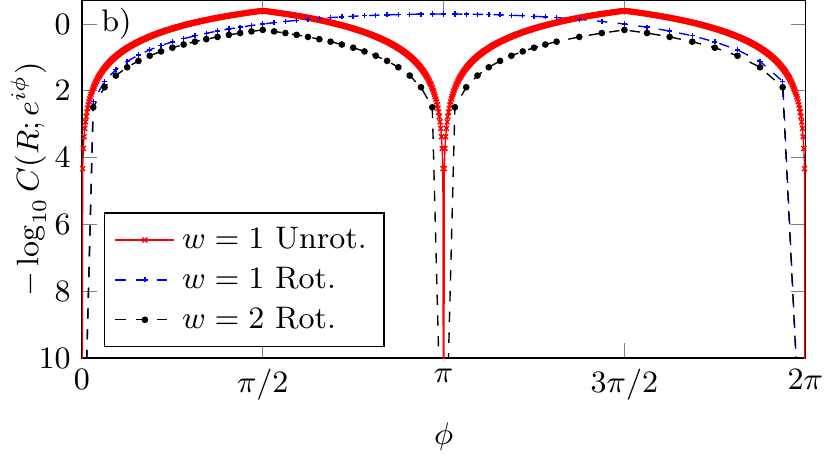}{
  \tikzsetnextfilename{RotatedZQDICostData}
  \setlength\figureheight{.42\columnwidth} 
  \setlength\figurewidth{.85\columnwidth}
  \input{graphs/RotatedTC.tikz}}
  \caption{Subfigures a) and b) show the cost of ribbons with supports oriented as shown in \Fref{fig:RotatedRibbons} for two models: a) the quantum compass model and b) the QDI$_2$ model. The cost is parameterized with $\eta = \mathrm{e}^{i \phi}$, and all ribbons were found using bond dimension 5. \\
a) Cost of ribbons in the quantum compass model of \Eref{eqn:QCmodel}. Because ribbons in this model are sensitive to their support, we conclude that they do not represent genuine topological order.\\
b) Cost of ribbons in the QDI$_2$ model of \Eref{eqn:QDI} using $\{J,h,\lambda\}=\{1,0,0\}$. Note that the unit cell in the rotated case is $4w-2$ rather than $2w$ for convenience. Low cost ribbons can be found on both unrotated and rotated supports indicating genuine topological order.}\label{fig:RotatedRibbonsCostData}
\end{figure}

In \Fref{fig:RotatedRibbonsCostData} we show the cost for ribbons obtained in these two models where the vertical ribbons are fixed to be $\prod Z_j$ and a bond dimension $\leq 5$ ribbon is optimized with both horizontal and rotated support. As discussed above, zero cost ribbons can be obtained for both models in the unrotated case when $\eta=-1$. Thus we see that each model supports a topologically encoded qubit. Once we rotate the ribbons, we find that in both cases the ribbon disappears when we restrict to $w=1$. In the QDI$_2$, the ribbon is recovered once we allow the support to grow to $w=2$, however in the QC model there is no indication of a good ribbon existing even when we allow ribbons of width $3$. This indicates that the low cost ribbons in the QC model are anomalies associated to lattice symmetries instead of genuine signatures of topological order.

\subsection{Kitaev's Honeycomb Model}
The honeycomb model\cite{Kitaev2006} is a frustrated spin model on the honeycomb lattice with Hamiltonian
\begin{align}
H&=-J_X\hspace{-3eX}\sum_{i,j\in X\,\text{links}}\hspace{-2.5eX}X_iX_j-J_Y\hspace{-3eX} \sum_{i,j\in Y\,\text{links}}\hspace{-2.5eX} Y_iY_j-J_Z\hspace{-3eX}\sum_{i,j\in Z\,\text{links}}\hspace{-2.5eX} Z_iZ_j\,,\label{eqn:HoneycombH}
\end{align} 
where the $X$, $Y$, and $Z$ links refer to the three orientations of edges on the lattice. For $0<J_X+J_Y<J_Z$, this model supports a phase with the same order as the $\ZZ_2$ quantum double model.\cite{Kitaev2006} Writing a pair of anticommuting logical operators for this phase is a nontrivial task however, since they cannot both be symmetries of the Hamiltonian and will hence not have completely localized support. We can instead attempt to describe them with ribbons having some width $w$. 

\begin{figure}
\centering
  \includeTikz{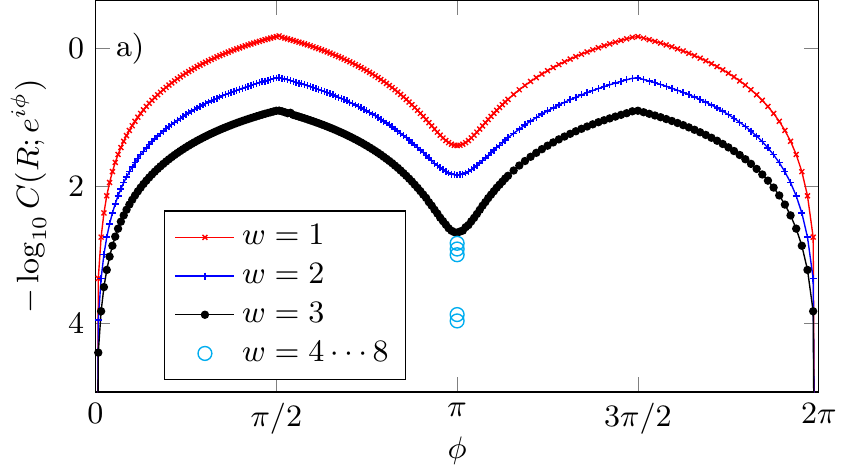}{
  \tikzsetnextfilename{HoneycombWidth}
      \setlength\figureheight{.46\linewidth} 
      \setlength\figurewidth{.85\linewidth} 
 \input{graphs/HoneycombWidth.tikz}}
     \\
    \includeTikz{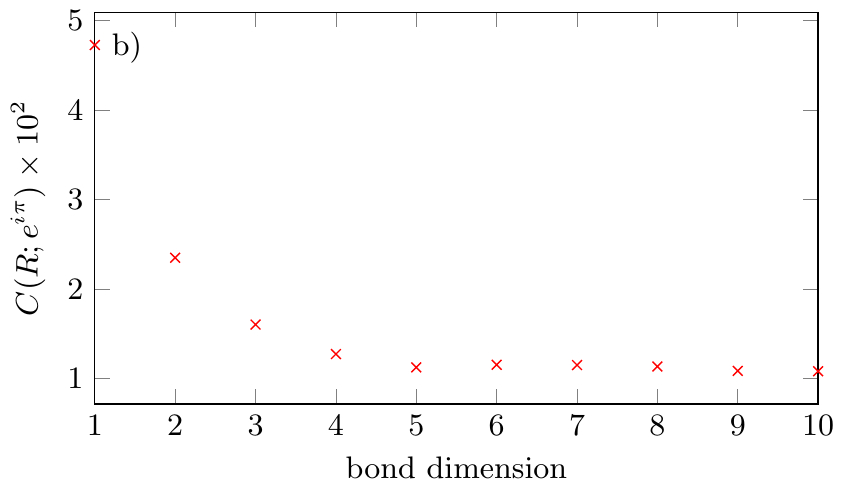}{
      \tikzsetnextfilename{Honeycomb_bd}
      \setlength\figureheight{.46\linewidth} 
      \setlength\figurewidth{.85\linewidth} 
      \input{graphs/Honeycomb_bd.tikz}}
  \\
    \includeTikz{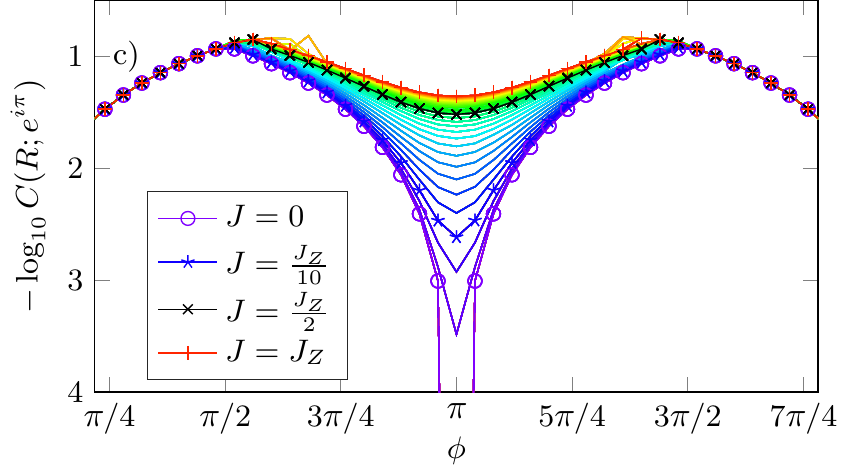}{
    \tikzsetnextfilename{IncreasingJx}
      \setlength\figureheight{.46\linewidth} 
      \setlength\figurewidth{.85\linewidth} 
    \input{graphs/IncreasingJx_new.tikz}}
  \caption{Cost of ribbons in the Honeycomb model. Strong indications of the $\ZZ_2$ topological order known to be present in this phase are visible. As expected, the cost decreases as the allowed width or bond dimension is increased, however the cost rapidly saturates with increasing bond dimension.\\
  a)  Cost for a bond dimension 5 ribbon on the honeycomb model of \Eref{eqn:HoneycombH} with $J_X=J_Y=J_Z/10$ using $\eta = \mathrm{e}^{i \phi}$. A second ribbon can be obtained analytically, and was fixed as a product of $Z$s along the $X$ and $Y$ links of the lattice. The dip at $\phi=\pi$ is characteristic of $\ZZ_2$ order, and indicates the topological nature of this phase.\\
  b) Cost for a width 2 ribbon at $\eta=-1$ on the Honeycomb model with $J_X=J_Y=J_Z/10$ as a function of bond dimension. The cost rapidly saturates to a minimum, justifying our assumption that only small bond dimension is required.\\
  c) As we increase the strength of $J=J_X=J_Y$, we eventually cross a phase transition to a gapless phase at $J=J_Z/2$ (indicated by $\times$). Approaching this point, the signal of $Z_2$ topological order essentially vanishes. Here the strength of $J$ is increased in steps of $1/30$ from $J=0$ ($\circ$) to $J=1$ ($+$). Note that the line marked by $\star$ corresponds to plots $a,b$ and a bond dimension of $4$ is used for a width $3$ ribbon.}\label{fig:HoneycombCostData}
\end{figure}

In \Fref{fig:HoneycombCostData}, we investigate ribbon operators for the honeycomb model. As before, we observe the characteristic dip at $\eta=-1$ signalling a topologically ordered phase (Fig.~\hyperref[fig:HoneycombCostData]{\ref*{fig:HoneycombCostData}a}). We also observe the effect of increasing the allowed ribbon width in Fig.~\hyperref[fig:HoneycombCostData]{\ref*{fig:HoneycombCostData}a}. As expected the cost goes down as the width increases, although this is not a smooth decrease. We suggest that this is due to the nature of the widening. Each time the width is increased by three, an additional plaquette lies completely within the support of $R$. These plaquettes commute with the Hamiltonian, and so correspond to conserved quantities. In the perturbative regime, these plaquettes become those in the toric code model, which are associated with the location of anyonic charge. We note that the correlation length for $J_X=J_Y=J_Z/10$ is approximately $.21$,\cite{Yang2008} so all of the widths correspond to several correlation lengths.

We also investigate how restricting the bond dimensions affects the cost. In Fig.~\hyperref[fig:HoneycombCostData]{\ref*{fig:HoneycombCostData}b}, we use a fixed width (3) and a fixed strength of $J_X=J_Y=J_Z/10$. At this point, we see that the cost rapidly saturates to a minimum for a small bond dimension of 4-5. This justifies the relatively small bond dimensions used in this work. 

Finally, we examine the effect of increasing the strength of $J=J_X=J_Y$ in Fig.~\hyperref[fig:HoneycombCostData]{\ref*{fig:HoneycombCostData}c}. When $J_X=J_Y=J_Z/2$, the model undergoes a transition from a phase supporting $\ZZ_2$ topological order to a gapless phase. The latter supports Ising anyons if a time reversal symmetry breaking term is added.\cite{Kitaev2006} Although we do not see a strong signal, such as a discontinuity, at this point, we see that the dip associated with $\eta=-1$ becomes very shallow. We expect the width of the required ribbons to be comparable to the correlation length, which we expect to increase as we move out of the perturbative regime. Thus the width $3$ ribbon used here is expected to be inadequate for large $J_X$ even within the phase. It remains to be seen whether ribbon operators can be used to identify phase transitions, possibly using a scaling analysis of ribbon width and bond dimension.

\section{Ribbons As Logical Operators}\label{S:LogicalOperators}
%------------------------------------------------------------------------------------------------------------%

One of the motivations for the ribbon operator methods is an application to quantum error correcting codes. In this section we show how nontrivial ribbon operators with sufficiently low cost can be used to certify the existence of an approximate topological quantum error correcting code in the ground space of the model. The ribbon operators function as approximate logical operators for these approximate codes. For gapped phases, this will also provide a natural justification for the method, although to apply this justification to the numerical results of \Sref{S:Results} would require similar proofs with weaker assumptions than we are currently able to make here.

Let $H$ be a Hamiltonian with gap $\Delta$ and suppose that for some $\epsilon>0$, $\delta>0$ and $\eta$, there exist ribbon operators $L$ and $R$ such that $\opnorm{\comm{H,R}}\leq\epsilon$, $\opnorm{\comm{H,L}}\leq\epsilon$, and $\bigl\|{\comm{R,L}_\eta}\bigr\|_{\text{op}}\leq \delta$. Here $\opnorm{R}=\opnorm{L}=1$, we restrict to hermitian ribbons, and we let $\opnorm{\cdot}$ denote the operator norm. Let $\ket{g}$ be some ground state of $H$ with energy $E_0$. Then the expected energy of $\ket{h}=R\ket{g}$ is
\begin{align}
\bra{g}R^\dagger H R\ket{g}&=\bra{g}R^\dagger RH\ket{g}+\bra{g}R^\dagger\comm{H,R}\ket{g}\nonumber\\
&\leq E_0+\epsilon.
\end{align}
Thus, $\ket{h}$ is a low energy state, and we can interpret $R$ as a generator for a unitary approximate logical operator. 

It is important to check that $R$ is not mapping out of the ground space, so we need to ensure that there is a large overlap.
Let $\Pi$ denote the ground space projector for $H$. By assumption the model has a spectral gap $\Delta$, so we find that
\begin{align}
E_0+\epsilon&\geq \bra{h}H\ket{h}\nonumber\\
&\geq E_0\bra{h}\Pi\ket{h}+(E_0+\Delta)\bra{h}\bm{1}-\Pi\ket{h}\,.
\end{align}
Rearranging this, and noting that $E_0$ can always be set to $0$, we obtain
\begin{align}
\frac{\bra{h}\Pi\ket{h}}{\braket{h}{h}}&\geq 1-\frac{\epsilon}{\braket{h}{h}\Delta}.
\end{align}
Thus, the action of the ribbon operators approximately preserves the ground space as long as they approximately preserve the norm of the individual ground states ($\braket{h}{h}\approx 1$). If this condition does not hold, $R$ is annihilating the ground space. 

Given an almost commuting pair of hermitian operators $R$ and $H$, we can perturb to an exactly commuting hermitian pair $\tilde{R}$ and $\tilde{H}$ such that the following holds:\cite{MR1424963,Hastings2009} $\bigl\|{R-\tilde{R}}\bigr\|_{\text{op}}\leq \gamma(\epsilon)$ and $\bigl\|{H-\tilde{H}}\bigr\|_{\text{op}}\leq \gamma(\epsilon)$, where $\gamma(\epsilon)$ can be taken to be at most $\epsilon^{1/30}$. It is convenient to work with $\tilde{H}$ and $\tilde{R}$ since they can be simultaneously exactly diagonalized. We need to check that the assumed twisted commutation relation between $R$ and $L$ is approximately maintained when we consider $\tilde{R}$ and $L$. 
\begin{align}
\delta&\geq \opnorm{RL-\eta LR} \nonumber \\
&=\bigl\|{(R-\tilde{R}+\tilde{R})L-\eta L(R-\tilde{R}+\tilde{R})}\bigr\|_{\text{op}} \nonumber \\
&\geq \bigl\|{\tilde{R}L-\eta L\tilde{R}}\bigr\|_{\text{op}}\!\!\!\!-\bigl\|{(R-\tilde{R})L-\eta L(R-\tilde{R})}\bigr\|_{\text{op}}.
\end{align}
Rearranging and using standard inequalities, we find
\begin{align}
\bigl\|{\tilde{R}L-\eta L\tilde{R}}\bigr\|_{\text{op}} \leq \delta+(1+|\eta|)\gamma,
\end{align}
and so the twisted commutator is approximately preserved. The same argument shows that $L$ approximately commutes with $\tilde{H}$ and thus approximately preserves its ground space.

We now check that the action of $L$ on eigenstates of $\tilde{R}$ maps to a nearly orthogonal state and can therefore be used to manipulate information in the code space.
Since $\tilde{H}$ and $\tilde{R}$ exactly commute, we can find a joint eigenstate $\ket{\tilde{g}}$ such that $\tilde{R}\ket{\tilde{g}}=\tilde{g}\ket{\tilde{g}}$ and $\tilde{H}\ket{\tilde{g}}=E_0\ket{\tilde{g}}$, then
\begin{align}
\delta&\geq|\bra{\tilde{g}}RL-\eta LR\ket{\tilde{g}}|\,,
\end{align}
and therefore
\begin{align}
|\bra{\tilde{g}}L\ket{\tilde{g}}|&\leq\frac{\delta}{|\tilde{g}||1-\eta|}\,.
\end{align}
We have already assumed that $R$ maps a ground state to an approximately normalized state, so $|\tilde{g}|\approx 1$. Therefore, the action of $L$ maps $\ket{\tilde{g}}$ to an approximately orthogonal state.

Thus, even approximate ribbon operators preserve the ground space (or annihilate it). Together, these results provide a certificate of topological order, at least when the strong assumptions of the derivation are met. In addition to providing such a certificate, this also shows that the method of ribbon operators can be used to obtain approximate logical operators even if the topological phase is already known.  The ribbons can be used to enact logical operations on encoded qudits even when the underlying model is not an exact topological quantum code, such as in the case of the  honeycomb model. 

Recent work has investigated the possibility of certifying degeneracy in the ground space of a Hamiltonian using ``approximate symmetries''.\cite{Chubb2016} In the case of unitary ribbons and $\eta=\exp(i \theta)$, the authors showed that if $\norm{\comm{H,R}}$, $\norm{\comm{H,L}}$, and $\bigl\|{\comm{R,L}_\eta}\bigr\|$ are sufficiently small, then the degeneracy of the ground space can indeed be certified as larger than some integer specified by $\theta$. In the nontrivial case (i.e.\ ground space degeneracy larger than 1), the operators act as approximate logical operators when restricted to the ground space. One could utilize this result, along with the numerical procedures outlined in this work, to \emph{prove} topological degeneracy by investigating the behavior of the cost function on different manifolds.

\section{Summary and Outlook}\label{S:future}
%------------------------------------------------------------------------------------------------------------%

We have introduced the method of ribbon operators for detecting topological order. By identifying features expected in topologically ordered spin models, we have defined a cost function which quantifies the extent to which operators on the lattice realize these features.

Using a variational minimization algorithm on this cost function over the space of matrix product operators, we have demonstrated that this method can distinguish nontrivial topological order from non-topological phases in various models, both integrable and nonintegrable. We have also shown how, with additional assumptions, ribbon operators can be used as approximate logical operators in topological quantum error correcting codes, which provides a specific sense in which they can certify the presence of topological order.

The most obvious open question is to extend these methods to other models, including those with nonabelian topological order. These nonabelian models are particularly interesting from a quantum information perspective as they can be utilized for universal quantum computation.\cite{Kitaev2003,Nayak2008} Another open question is whether this method can be extended to detect symmetry-protected or symmetry-enriched topological order, but this would seem to require new ideas.

Even without generalizing the method to deal with nonabelian models, there are several natural ways in which our approach can be improved and extended. One could proceed by either changing the cost function, the variational class, or the optimization method. We begin by discussing a natural restriction on the allowed MPOs.

As shown in \Sref{S:QDIResults}, some topologically trivial models support zero-cost ribbon operators when the variational class is all MPOs of a given bond dimension. These operators corresponded to products of projectors onto local ground states. Although such operators have low cost, they are not good signals of topological order, since the $\eta$ for which $\comm{R,L}_\eta=0$ is not unique because $R$ and $L$ mutually annihilate. This behavior could be removed by insisting that the ribbon operators be unitary. In addition to removing these false signals, a unitary constraint would ensure that ribbon operators properly preserve the norm of states on which they act. As discussed in \Sref{S:LogicalOperators}, this property is important if the ribbons are to be interpreted as approximate logical operators for a quantum error correcting code.

A potential failure mode for this method is `high-temperature topological order'. Suppose there is some gapped state high up in the spectrum of the Hamiltonian, for example, at the top of the spectrum. It is conceivable that such a state could posess a `topological order' distinct from the low-energy space. In this case, the method may find signals of TO which are originating from the high-energy portion of the spectrum. Such a failure mode could be combated by incorporating a low-temperature thermal state into the cost function. This would bias the norm towards the low-energy space, without the need for computing the ground state itself. Note that low temperature thermal states have efficient PEPS descriptions,\cite{Hastings2006,Molnar2015} so the computational benefits of this method may remain.

One may also consider applying the method to seek out such high-energy topologically ordered subspaces in models with trivial ground spaces. This may provide an avenue towards interesting thermal physics with a topological flavor.

There are various alterations which can be made to the cost function defined in \Sref{S:CostFunction}. Recall that our cost function used the $\eta$-commutator to incorporate the topological data. This is associated with the $\mathcal{R}$ matrix of the anyon model. A natural replacement for, or addition to, this term would be a term involving the $\tilde{\mathcal{S}}$ matrix. It is an open question if the current method or any of these suggested generalizations can be used to obtain the complete $\mathcal{S}$ and $\mathcal{R}$ matrices of a model. Recall from \Sref{S:ribbons} that in the abelian case, there is a column permutation of $\mathcal{S}$ which cannot be fixed. This could be addressed by ensuring all ribbons lie on comparable supports, for example circles or L shaped regions.

As discussed in \Sref{S:ribbons}, insisting that a ribbon $R$ commute with the Hamiltonian on all eigenspaces (e.g.\ using the Frobenius norm $\norm{\comm{R,H}}$) is probably too strong to identify the topological order present in all models. One usually discusses topological order only with respect to the low-energy sector of the model. We propose that rather than taking the norm of the commutator, a variational low-energy state, for example in PEPS\cite{Schuch2010} or MERA\cite{Vidal2008} form, could be used to supplement the algorithm. Alternatively, one could use a low-energy thermal state as discussed above. Rather than evaluating norms, one could evaluate expectation values on these low energy states. This bypasses problems associated with finding ground states of 2D Hamiltonians, since the state need only be low energy and supported on a strip only just wider than the ribbon itself. Access to a state may be required for the extension to nonabelian models, where the excited spectrum is expected to be more exotic than in the abelian case and reflect the structure of the fusion space of the underlying anyons.

It would be interesting to apply the method to models where the topological order is still debated, for example the Heisenberg model on the Kagome lattice. When our cost function is encoded into an MPO for minimization using DMRG, it has a large bond dimensions since it is quadratic in the Hamiltonian of the model. This makes it computationally expensive to minimize. To approach these more complex models, it may be necessary to make use of properties of this MPO, such as sparsity, to reduce this cost. Alternatively, one could modify the cost function to reduce the bond dimension, or use another method to minimize it.

The assumptions used in \Sref{S:LogicalOperators} are stronger than the constraints imposed in our numerical work. Understanding how to weaken these assumptions would provide additional justification for this method, and may provide insight into possible additional constraints which a ribbon operator should obey to ensure it is truly certifying topological order. It would also be interesting to blend our algorithm with the results of \cite{Chubb2016} with a view to certifying topological degeneracy.

\vspace*{-3mm}
\acknowledgments
\vspace*{-3mm}
We thank Matthias Bal, Parsa Bonderson, Christopher Chubb, Andrew Doherty, Christopher Granade, Jeongwan Haah, Jutho Haegeman, Robert Pfeifer, Sam Roberts, Norbert Schuch, Tom Stace, Frank Verstraete and Dominic Williamson for useful and enlightening discussions. We also thank the anonymous referee who pointed out the potential failure mode that may result from a model with high-energy topologically ordered subspaces. We acknowledge support from the Australian Research Council via the Centre of Excellence in Engineered Quantum Systems (EQuS), project number CE110001013. STF also acknowledges support from an Australian Research Council Future Fellowship FT130101744. This work was initiated while DP was on sabbatical leave at The University of Sydney, he acknowledges their hospitality and their International Research Collaboration Award for partial support. 

%------------------------------------------------------------------------------------------------------------%
%------------------------------------------------------------------------------------------------------------%
%
%------------------------------------------------------------------------------------------------------------%
\onecolumngrid
\appendix
\section{$\mathbb{Z}_2$ Quantum Double}\label{S:costfunc}
The cost function for a pair of width 1 ribbon operators in the $\ZZ_2$ quantum double model defined in Eq.~18 of the main document (with $(J,h,\lambda)=(1,0,0)$) can easily be written. Define the lattice as in Fig.~1 of the main document, and let the strings $R$ and $L$ have supports on the horizontal and vertical strips respectively. Beginning with ribbon $R$, we see that there are no Hamiltonian terms contained completely within the support. Thus the first term in the cost vanishes. The same happens for ribbon $L$. 

The next term concerns those Hamiltonian terms which cross the boundary. There are two classes, those with support on a single spin inside $R$ and those supported on a pair of spins in $R$. Each dark plaquette (\raisebox{-1pt}{\includeTikz{DarkPlaquette}{\tikzsetnextfilename{DarkPlaquette}\tikz[scale=.25,baseline=-2pt] \filldraw[fill=black!30](-.5,0)--(0,.5)--(.5,0)--(0,-.5)--(-.5,0);}}\!\!) contributes to a term in the first class, with each spin $j$ in $R$ being touched by a pair of such plaquettes. Therefore at this stage we have an initial cost function $C_0$ for the commutator part given by
\begin{align}
C_0(R)&=\left(2\sum_{j=1}^N\norm{\comm{R,X_j}}^2\right),
\end{align}
where $N$ is the number of spins on which $R$ is supported. Each light plaquette (\raisebox{-1pt}{\includeTikz{LightPlaquette}{\tikzsetnextfilename{LightPlaquette}\tikz[scale=.25,baseline=-2pt] \filldraw[fill=black!00](-.5,0)--(0,.5)--(.5,0)--(0,-.5)--(-.5,0);}}\!\!) contributes a two-spin term, so the costs become
\begin{align}
C_0(R)&=\left(2\sum_{j=1}^N\norm{\comm{R,X_j}}^2+\sum_{j=1}^N\norm{\comm{R,Z_jZ_{j+1}}}^2\right).
\end{align}

For convenience, assume two-site translationally invariant product operators for both $R=\cdots abab\cdots$ and $L=\cdots cdcd\cdots$.  This cost function becomes
\begin{align}
C_0(R)=&2\frac{N}{2}\norm{\comm{a,X}}^2+2\frac{N}{2}\norm{\comm{b,X}}^2+\frac{N}{2}\norm{\comm{ab,ZZ}}^2+\frac{N}{2}\norm{\comm{ba,ZZ}}^2,
\end{align}
since $N/2$ of the sites of $R$ support operator $a$ and $N/2$ sites support operator $b$.

There are no more terms contributed by the commutator with the Hamiltonian. It remains to include the twisted commutator terms. The ribbon $L$ can intersect $R$ with either a $c$ site or a $d$ site. Recall that we require that all translations of $L$ are considered, so we obtain
\begin{align}
C_1(R;\eta)=\sum_j \norm{\comm{r_j,c}_\eta}^2+\sum_j \norm{\comm{r_j,d}_\eta}^2,
\end{align}
where $r_j$ is the operator at site $j$ of ribbon $R$. Notice that each $r_j$ is required to $\eta-$commute with both $c$ and $d$. In this way, the insensitivity of this term to the particular support of $L$ is ensured. This term gives
\begin{align}
C_1(R;\eta)&=\frac{N}{2}\norm{\comm{a,c}_\eta}^2+\frac{N}{2}\norm{\comm{b,c}_\eta}^2+\frac{N}{2}\norm{\comm{a,d}_\eta}^2+\frac{N}{2}\norm{\comm{b,d}_\eta}^2.
\end{align}
Combining these two costs, we obtain the total cost function for ribbon $R$
\begin{align}
C(R;\eta)=&\frac{1}{N}\left(C_0(R)+C_1(R;\eta)\right)\\
=&\norm{\comm{c,Z}}^2+\norm{\comm{d,Z}}^2+\frac{1}{2}\norm{\comm{cd,XX}}^2+\frac{1}{2}\norm{\comm{dc,XX}}^2\nonumber\\
&+\frac{1}{2}\norm{\comm{a,c}_\eta}^2+\frac{1}{2}\norm{\comm{b,c}_\eta}^2+\frac{1}{2}\norm{\comm{a,d}_\eta}^2+\frac{1}{2}\norm{\comm{b,d}_\eta}^2.
\end{align}
The cost for ribbon $L$ can be obtained analogously, giving
\begin{align}
C(L;\eta)=&\norm{\comm{c,Z}}^2+\norm{\comm{d,Z}}^2+\frac{1}{2}\norm{\comm{cd,XX}}^2+\frac{1}{2}\norm{\comm{dc,XX}}^2\nonumber\\
&+\frac{1}{2}\norm{\comm{a,c}_\eta}^2+\frac{1}{2}\norm{\comm{b,c}_\eta}^2+\frac{1}{2}\norm{\comm{a,d}_\eta}^2+\frac{1}{2}\norm{\comm{b,d}_\eta}^2.
\end{align}
Notice that $C_1(R;\eta)=C_1(L;\eta)$. Care should be taken with this term since $\comm{\,\cdot\,,\cdot\,}_\eta$ is not symmetric in its arguments.

It is easy to check that, as expected, these cost functions are both zero when we set $a=b=X$, $c=d=Z$ and $\eta=-1$, corresponding to the known string operators for this model. 
\end{document}